\documentclass[journal]{IEEEtran}
\usepackage{amsmath,bm,mathtools,commath}
\usepackage[scr=rsfs]{mathalfa}
\usepackage{amssymb}  % for math spacing
\usepackage{graphicx}
\usepackage{epsfig}
\usepackage{epstopdf}
\usepackage{verbatim}
\usepackage{multirow}
\usepackage{hhline}
\usepackage{url}
\usepackage{array}
\usepackage{makeidx}         % allows index generation
\usepackage[bottom]{footmisc}% places footnotes at page bottom
\usepackage{paralist}
\usepackage{caption}
\usepackage{subfig}
\usepackage{amsthm}

\usepackage{epstopdf}

\theoremstyle{plain}
\newtheorem{assumption}{Assumption}[]
\theoremstyle{plain}
\newtheorem{observation}{Observation}[]
\theoremstyle{Definition}

\theoremstyle{Theorem}

\allowdisplaybreaks
\begin{document}
	%
	% paper title
	% Titles are generally capitalized except for words such as a, an, and, as,
	% at, but, by, for, in, nor, of, on, or, the, to and up, which are usually
	% not capitalized unless they are the first or last word of the title.
	% Linebreaks \\ can be used within to get better formatting as desired.
	% Do not put math or special symbols in the title.
	\title{A Library of Second-Order Models \\for Synchronous Machines}
	%
	%
	% author names and IEEE memberships
	% note positions of commas and nonbreaking spaces ( ~ ) LaTeX will not break
	% a structure at a ~ so this keeps an author's name from being broken across
	% two lines.
	% use \thanks{} to gain access to the first footnote area
	% a separate \thanks must be used for each paragraph as LaTeX2e's \thanks
	% was not built to handle multiple paragraphs
	%

	\author{Olaoluwapo~Ajala,~\IEEEmembership{Student Member,~IEEE,}
		Alejandro~Dom\'{i}nguez-Garc\'{i}a,~\IEEEmembership{Member,~IEEE,}
		Peter~Sauer,~\IEEEmembership{Life~Fellow,~IEEE,}
		and~Daniel~Liberzon,~\IEEEmembership{Fellow,~IEEE}% <-this % stops a space
		\thanks{The authors are with the Department of Electrical and Computer Engineering, University of Illinois at Urbana-Champaign, Urbana, IL 61801 USA. E-mail:\{ooajala2, aledan, psauer, liberzon\}@ILLINOIS.EDU.}}% <-this % stops a space
	\maketitle
	
	% As a general rule, do not put math, special symbols or citations
	% in the abstract or keywords.
	\begin{abstract}
		This paper presents a library of second-order models for synchronous machines that can be utilized in power system dynamic performance analysis and control design tasks. The models have a similar structure to the classical model in that they consist of two dynamic states, the power angle and the angular speed. However, unlike the classical model, the models find applications beyond first swing stability analysis; for example, they can also be utilized in transient stability studies. The models are developed through a systematic reduction of a nineteenth-order model, using singular perturbation techniques, and they are validated by comparing their voltage, frequency, and phase profiles with that of the high-order model and that of the classical model.
	\end{abstract}
	
	% Note that keywords are not normally used for peerreview papers.
	
	% For peer review papers, you can put extra information on the cover
	% page as needed:
	% \ifCLASSOPTIONpeerreview
	% \begin{center} \bfseries EDICS Category: 3-BBND \end{center}
	% \fi
	%
	% For peerreview papers, this IEEEtran command inserts a page break and
	% creates the second title. It will be ignored for other modes.
	\IEEEpeerreviewmaketitle

	\section{Introduction}
	% The very first letter is a 2 line initial drop letter followed
	% by the rest of the first word in caps.
	% 
	% form to use if the first word consists of a single letter:
	% \IEEEPARstart{A}{demo} file is ....
	% 
	% form to use if you need the single drop letter followed by
	% normal text (unknown if ever used by the IEEE):
	% \IEEEPARstart{A}{}demo file is ....
	% 
	% Some journals put the first two words in caps:
	% \IEEEPARstart{T}{his demo} file is ....
	% 
	% Here we have the typical use of a "T" for an initial drop letter
	% and "HIS" in caps to complete the first word.
	\IEEEPARstart{D}{ynamic} models of synchronous machines find applications in power system analysis, control design tasks, and education, with each application requiring models that capture dynamical phenomena relevant to the intended use. This has led to the proliferation of synchronous machine models in the literature \cite{Kundur1994,sauer2006power,KrauseWasynczuk2013,Wang2007}, with varying degrees of complexity, computational cost, and state-space dimension. One such model is the so-called classical model advocated in \cite{Crary1947} and \cite{Kimbark1956}, a second-order dynamic model that captures the dynamics of the machines phase and angular speed.
	
	Analytically, the classical model is the simplest synchronous machine dynamical model, but it has certain limitations that restrict its applications to first swing stability analysis, i.e. stability analysis for the first second \cite{pai1989energy,anderson2003power,classicalmodel2015}. As a result, if we consider that a power system may be stable in the first swing but unstable in subsequent swings, it is clear that the classical model, though simple, is unreliable for power system tasks extending beyond a one second time interval. For example, the design of a generator synchronization scheme requires a model that captures dynamics of the generator phase, frequency and voltage magnitude over the entire synchronization period. A second-order model such as the classical model should suffice, but the first swing stability constraint could make it inapplicable if the synchronization period exceeds one second. On the other hand, while existing high-order models, such as the two-axis model and the one-axis model \cite{sauer2006power,Weckesser2013}, are clearly more accurate and therefore very useful for power system simulation, they are also significantly more detailed and computationally expensive. Consequently, the high-order models are, in general, analytically intractable for such control design tasks. There is therefore a need to develop models that possess the simplicity of the classical model, but also the temporal breadth that it lacks.
	
	%In such applications, power system engineers typically utilize more detailed high-order models, however these models may prove too complex to be useful for some controller design and synchronization analysis tasks \cite{sauer2006power}.
	The main contribution of this paper is the development of second-order synchronous machine models that, when compared to the classical model, have the same state-space dimension, are significantly more accurate over a long time interval, and are useful for a broader range of applications. Using singular perturbation analysis as our main tool \cite{sauer2006power,kokotovicSingular,Khalil2013,joechow1982,SauerAhmedKokotovic1988}, the second-order models presented in this paper are derived by
	\begin{inparaenum}[(i)]
		\item identifying the fastest dynamic states in a high-order model;
		\item  developing approximate manifold equations for them, which are algebraic equations; and
		\item replacing the differential equations for these states with the algebraic counterparts.
	\end{inparaenum}
	
	Our approach to developing the proposed machine models is based on the developments in \cite{sauer2006power,SauerAhmedKokotovic1988,SauerKokotovic1989}, where zero-order and first-order approximations of manifolds for fast dynamic states are used to develop reduced-order models. In \cite{SauerAhmedKokotovic1988,SauerKokotovic1989}, the use of integral manifolds for model order reduction is introduced with some applications presented, and in \cite{sauer2006power}, the technique is used to develop the two-axis model, the one-axis model, and the classical model.
	
	The remainder of the paper is organized as follows. In Section~\ref{sec:prelim}, we present a synchronous machine and a high-order model that is adopted as the starting point for the development of our reduced order models; we also discuss the classical model. In Section~\ref{sec:ROm}, we develop a library of second-order models from the high-order model, using singular perturbation analysis. Finally, in Section~\ref{sec:simRes} we validate the second-order models developed, using numerical examples, and in Section~\ref{sec:conclusion} we comment on implications of the presented results.
	% You must have at least 2 lines in the paragraph with the drop letter
	% (should never be an issue)
	
	\section{Preliminaries}\label{sec:prelim}
	We begin this section by presenting the high-order model of a synchronous machine adopted in this work. In addition, the time-scale properties of the model are discussed. Afterwards, we introduce the so-called classical model and describe how it can be developed from the high-order model.
	\subsection{High-Order Synchronous Machine Model}\label{sec:HOm}
	The high-order synchronous machine model we describe in this section is based on the developments in \cite{sauer2006power,KrauseWasynczuk2013}. The components included in the model are: \begin{inparaenum}[(i)]
		\item three damper windings,
		\item a wound-rotor synchronous machine,
		\item an IEEE type DC1A excitation system \cite{IEEE_ExcMod}, and
		\item a Woodward diesel governor (DEGOV1) \cite{Degov1MOD}, coupled to a diesel engine, which acts as the prime mover.
	\end{inparaenum}
	Next, we provide mathematical expressions that describe the dynamic behavior of these components. [Note that the model is presented utilizing the $qd0$ transformation, with all parameters and variables scaled, and normalized using the per-unit system].
	\begin{assumption}\label{assum:machine}
		The synchronous machine is connected to an electrical network bus through a short transmission line.
	\end{assumption}
	\subsubsection{Damper windings model}\label{sec:damper}
	Let $\Phi_{q_2}(t)$ and $E_{d'}(t)$ denote the flux linkages of two damper windings aligned with the quadrature axis ($q$-axis) of the synchronous machine, let $\Phi_{d_1}(t)$ and $E_{q'}(t)$ denote the flux linkages of a damper winding and a field winding, respectively, aligned with the direct axis ($d$-axis) of the synchronous machine, and let $I_{q}$ and $I_{d}$ denote the $q$-axis and $d$-axis components of the stator output current, respectively. Then, the damper winding dynamics can be described as follows:
	\begin{gather} 
	\begin{split}
	\tau_{q''} \dot{\Phi}_{q_2} =& -\Phi_{q_2} -\left(X_{q'}-X_{k}\right)I_q - E_{d'},\\
	\tau_{d''} \dot{\Phi}_{d_1} =& -\Phi_{d_1} - \left(X_{d'}-X_{k}\right)I_d + E_{q'}(t),
	\end{split}\label{eqn:fastdamp}
	\end{gather}
	and
	\begin{align}
	\tau_{q'} \dot{E}_{d'} =& - E_{d'}+\left(X_q-X_{q'}\right)\left(I_q-\frac{X_{q'}-X_{q''}}{(X_{q'}-X_{k})^2}\left(\Phi_{q_2}\right.\right.\nonumber\\&\left.\left.+(X_{q'}-X_{k})I_q-E_{d'}\right)\right),\label{eqn:slowdamp}
	\end{align}
	where $X_{k}$ denotes the machine leakage reactance, $X_q$ denotes the machine stator reactance, $X_{q'}$ and $X_{d'}$ denote machine transient reactances, and $X_{q''}$ denotes the machine sub-transient reactance and $\tau_{q''}=\frac{1}{\omega_0R_{q_2}}\left(X_{kq_2}+X_{mq}\right)$, $\tau_{d''}=\frac{1}{\omega_0R_{d_1}}\left(X_{k_1}+\frac{X_{md}X_{kf}}{X_{md}+X_{kf}}\right)$ and $\tau_{q'}=\frac{X_{kq_1}+X_{mq}}{\omega_0R_{q_1}}$ are time constants, with $X_{kq_2}$, $X_{kd_1}$, $X_{kf}$, $X_{kq_1}$ denoting leakage reactances, $X_{mq}$ and $X_{md}$ denoting mutual reactances, and $R_{q_2}$, $R_{d_1}$, $R_{q_1}$ denoting winding resistances.
	
	\subsubsection{Stator windings and network model}\label{sec:stator}
	Let $\Phi_{q}^{(s)}(t)$ and $\Phi_{d}^{(s)}(t)$ denote the $q$-axis and $d$-axis components of flux linkages for the stator windings, respectively, let {\begin{math}
		\Phi_{q}^{(e)}(t)= -X^{(e)}I_q,\; \text{and}\; \Phi_{d}^{(e)}(t)= -X^{(e)}I_d
		\end{math}} denote the $q$-axis and $d$-axis components of flux linkages for the electrical line, respectively, let ${\omega}^{(s)}(t)$ denote the machine angular speed, in electrical radians per second, and let ${\delta}^{(s)}(t)$ denote the power angle of the synchronous machine in electrical radians. At the electrical network bus, let $V^{(l)}$ and $\delta^{(l)}$ denote the voltage magnitude, in per unit, and the voltage phase relative to a reference frame rotating at the nominal frequency, in electrical radians, respectively. Let \begin{math}
	V_{q}^{(l)}\coloneqq V^{(l)}\cos\left({\delta}^{(s)}-\delta^{(l)}\right),\; V_{d}^{(l)}\coloneqq V^{(l)}\sin\left({\delta}^{(s)}-\delta^{(l)}\right),\; \Phi_{q}(t)\coloneqq~\Phi_{q}^{(s)}(t)~+~\Phi_{q}^{(e)}(t),\; \Phi_{d}(t)\coloneqq\Phi_{d}^{(s)}(t)+\Phi_{d}^{(e)}(t)
	\end{math}. Then, the stator winding and network dynamics are described by:
	\begin{gather}
	\begin{aligned}
	\dot{{\delta}}^{(s)}=&\ {\omega}^{(s)}(t)-\omega_0,\\
	\frac{1}{\omega_0}\dot{\Phi}_{q} =& - \frac{{\omega}^{(s)}(t)}{\omega_0}\Phi_{d} + V_{q}^{(l)}+\left(R_{s}+R^{(e)}\right)I_{q},\\
	\frac{1}{\omega_0}\dot{\Phi}_{d} =&\ \frac{{\omega}^{(s)}(t)}{\omega_0}\Phi_{q} + V_{d}^{(l)}+\left(R_{s}+R^{(e)}\right)I_{d},\\
	\frac{1}{\omega_0}\dot{\Phi}_{q}^{(e)} =&\ R^{(e)}I_{q} - \frac{{\omega}^{(s)}(t)}{\omega_0}\Phi_{d}^{(e)} - V_{q}^{(s)} + V_{q}^{(l)},\\
	\frac{1}{\omega_0}\dot{\Phi}_{d}^{(e)} =&\ R^{(e)}I_{d} + \frac{{\omega}^{(s)}(t)}{\omega_0}\Phi_{q}^{(e)} - V_{d}^{(s)} + V_{d}^{(l)},\\
	\Phi_{q} =& -X_{q''}^{(e)}I_{q} + \frac{X_{q'}-X_{q''}}{X_{q'}-X_{k}}\Phi_{q_2}- \frac{X_{q''}-X_{k}}{X_{q'}-X_{k}}E_{d'},\\
	\Phi_{d} =& -X_{d''}^{(e)}I_{d} + \frac{X_{d'}-X_{d''}}{X_{d'}-X_{k}}\Phi_{d_1}+ \frac{X_{d''}-X_{k}}{X_{d'}-X_{k}}E_{q'}(t),
	\end{aligned}\label{eqn:stator_network}
	\end{gather}
	where \begin{math}X_{q''}^{(e)}\coloneqq X_{q''}+X^{(e)},\;\text{\normalsize and}\; X_{d''}^{(e)}\coloneqq X_{d''}+X^{(e)}, \end{math} $X^{(e)}$ denotes the per-phase line reactance, $X_{d''}$ denotes a machine sub-transient reactance, $R^{(e)}$ denotes the per-phase line resistance, $R_{s}$ denotes the per-phase stator resistance, and $\omega_0$ denotes the nominal frequency in electrical radians per second.
	
	\subsubsection{Excitation system model}\label{sec:voltage}
	Let $E_{f}(t)$ denote the output voltage of the machines excitation system, let $U_{f}(t)$ denote the exciter control input, let $\bar{U}_{f}(t)$ denote the rate feedback variable of the voltage regulator, and let $V^{(s)} \coloneqq \sqrt{\left(V_q^{(s)}\right)^2+\left(V_d^{(s)}\right)^2}$.
	\begin{assumption}\label{assum:saturation}
		The effects of magnetic saturation on the machines excitation system are negligible.
	\end{assumption}\noindent
	Then, the dynamics of the machines excitation system can be described as follows:
	\begin{gather}
	\begin{aligned}
	\tau_{d'} \dot{E}_{q'} =& - E_{q'}-\left(X_d-X_{d'}\right)\left(I_d-\frac{X_{d'}-X_{d''}}{(X_{d'}-X_{k})^2}\left(\Phi_{d_1}\right.\right.\\&\left.\left.+(X_{d'}-X_{k})I_d-E_{q'}\right)\right) + E_f,\\
	\tau_f \dot{E}_f =& -K_fE_f + U_f,\\
	\tau_u \dot{U}_f =& -U_f + K_u\bar{U}_{f}-\frac{K_u\bar{K}_u}{\bar{\tau}_u}E_f + K_u\left(V_r^{(s)}-V^{(s)}\right),\\
	\bar{\tau}_u \dot{\bar{U}}_{f} =& -\bar{U}_{f} + \frac{\bar{K}_u}{\bar{\tau}_u}E_f,
	\end{aligned}\label{eqn:excitation}
	\end{gather}
	where $V_r^{(s)}$ denotes the reference voltage magnitude, $\tau_{d'}=\frac{X_f}{\omega_0R_f}$, $\tau_f=\frac{L_f}{K_g}$, $K_f=\frac{\bar{R}_f}{K_g}$, $\bar{\tau}_u=\frac{L_{t}+L_{m}}{R_t}$, $\bar{K}_u=\frac{N_{t2}}{N_{t2}}\frac{L_{m}}{R_t}$, $X_d$ denotes the machine stator reactance, $\tau_u$ denotes the amplifier time constant, $K_u$ denotes the amplifier gain, $X_f$ denotes the field winding reactance, $R_f$ denotes the field winding resistance, $L_f$ denotes the unsaturated field inductance, $K_g$ denotes the slope of the unsaturated portion of the exciter saturation curve, $\bar{R}_f$ denotes the exciter circuit resistance, $L_{t}$ and $L_{m}$ denote series and magnetizing inductances of the stabilizing transformer, which is used to stabilize the excitation system through voltage feedback \cite{sauer2006power}, respectively, $R_t$ denotes the series resistance of a stabilizing transformer, and $\frac{N_{t2}}{N_{t1}}$ denotes the turns ratio of the stabilizing transformer.
	
	\subsubsection{Prime mover and speed governor model}
	Let $T_m(t)$ denote the mechanical torque output of the machine. For the speed governor system, let $P_{a_2}(t)$ denote the output of its actuator, with $\dot{P}_{a_1} =\ P_{a_2}(t)$, and let $P_{b_2}(t)$ denote the output of its electric control box, with $\dot{P}_{b_1} =\ P_{b_2}(t)$. Let $\dot{P}_{u} = P_{a_1}(t) + \tau_{4}P_{a_2}(t)$ denote the valve position of the diesel engine, which acts as the prime mover. Then, the speed control system of the synchronous machine can be expressed as follows:
	\begin{gather}
	\begin{split}
	M\dot{{\omega}}^{(s)} =&\ T_{m} - \Phi_{d}(t)I_{q} + \Phi_{q}(t)I_{d} - \tilde{D}_0{\omega}^{(s)},\\
	\tau_m\dot{T}_{m} =& -T_{m} + P_{u},\\
	\tau_{a_2}\dot{P}_{a_2} =& - \frac{1}{\tau_{5}+\tau_{6}}\left(P_{a_1}(t) - \kappa\left(P_{b_1}(t)+\tau_{3}P_{b_2}\right)\right) -P_{a_2},\\
	\tau_{2}\dot{P}_{b_2} =&\ \frac{1}{\tau_{1}}\left(\frac{1}{\bar{D}_{0}\omega_0}\left(P_{c} - P_{u}\right)-\frac{1}{\omega_0}\left({{\omega}}^{(s)}-\omega_0\right)\right)\\&-P_{b_2}- \frac{1}{\tau_{1}}P_{b_1}(t),
	\end{split}\label{eqn:governor_engine}
	\end{gather} 
	where $\tau_{2}$, $\tau_{3}$ , $\tau_{4}$, $\tau_{5}$ and $\tau_{6}$ denote time constants of the control system, $\tau_{a_2}=\frac{\tau_{5}\tau_{6}}{\tau_{5}+\tau_{6}}$, $\kappa$ denotes a controller gain for the actuator, $P_{c}$ denotes the power change setting of the machine, $M$ denotes the inertia of the machine, $\tilde{D}_0$ denotes the friction and windage damping coefficient of the machine, $\tau_m$ denotes the time constant of the engine, and $\bar{D}_{0}=\frac{1}{R_D\omega_0}$, with $R_D$ denoting the droop coefficient. [Note that for salient pole machines, $X_{q}=X_{q'}$, so that $E_{d'}(t)=0$, and for round-rotor machines, $X_{q}=X_{d}$].
	
	\subsection{High-Order Model Time-Scale Properties}\label{sec:observation}
	The following observations are based on standard parameter values obtained from synchronous machine models in \cite{Kundur1994,sauer2006power,KrauseWasynczuk2013,Degov1MOD}, and an eigenvalue analysis of these models.
	\begin{observation}\label{obs:FastStates}
		The dynamics of $\Phi_{q_2}$, $\Phi_{d_1}$, $E_{d'}$, $\Phi_{q}$, $\Phi_{d}$, $\Phi_{q}^{(e)}$, $\Phi_{d}^{(e)}$, $E_{q'}$, $E_{f}$, $U_{f}$, $\bar{U}_{f}$, $T_{m}$, $P_{u}$, $P_{a_2}$, $P_{b_2}$, $P_{a_1}$ and $P_{b_1}$, are much faster than those of ${\omega}^{(s)}$ and ${\delta}^{(s)}$.
	\end{observation}
	\begin{observation}\label{obs:SmallParam}
		For $\epsilon=0.1$ denoting a constant, the parameters $R_s$, $\tau_{q''}$, $\tau_{q'}$, $\frac{1}{\omega_0}$, $\tau_f$, $\tau_u$, $\bar{\tau}_u$, $\tau_m$, $\tau_{a_2}$, $\tau_{2}$, $\tau_{1}$, $(\tau_{5}+\tau_{6})$, $\frac{\tau_{5}\tau_{6}}{(\tau_{5}+\tau_{6})}$, $\frac{1}{\kappa R_D}$ are $\bm{\mathcal{O}}\left(\epsilon\right)$\footnote{Consider a positive constant $\epsilon$, where ${\epsilon}<1$, and a function $f(\epsilon)$, defined on some subset of the real numbers.  We write
			\begin{math}
			f(\epsilon)=\bm{\mathcal{O}}\left(\epsilon^i\right)
			\end{math} if and only if there exists a positive real number $k$, such that:
			\begin{math}
			\abs{f(\epsilon)}\leq k{\epsilon^i}, \text{ as }\epsilon\to0.
			\end{math}}.
	\end{observation}\noindent
	Based on these observations, the nineteenth-order machine model described by \eqref{eqn:fastdamp}--\eqref{eqn:governor_engine} can be expressed compactly as:
	\begin{equation}
	\begin{split}
	\dot{\bm{x}}(t)=&\ f\left(\bm{x}(t),\bm{z}(t),\epsilon\right),\qquad\bm{x}(0)=\bm{x}^0,\\
	\epsilon\dot{\bm{z}}(t)=&\ g\left(\bm{x}(t),\bm{z}(t),\epsilon\right),\qquad\,\bm{z}(0)=\bm{z}^0,
	\end{split}\label{eqn:compact15}
	\end{equation}
	where $\bm{x}(t) = \left[\begin{matrix}{\delta}^{(s)}&\omega^{(s)}\end{matrix}\right]^{\top}$, and $\bm{z}(t) = \left[\begin{matrix}\Phi_{q}&\Phi_{d}&E_{d'}\end{matrix}\right.\\\left.\begin{matrix}\Phi_{q_2}&\Phi_{d_1}&\Phi_{q}^{(e)}&\Phi_{d}^{(e)}&E_{q'}&E_{f}&U_{f}&\bar{U}_{f}&T_{m}&P_{u}\end{matrix}\right.\\\left.\begin{matrix}P_{a_2}&P_{b_2}&P_{a_1}&P_{b_1}\end{matrix}\right]^{\top}$. In the remainder of this paper, we refer to the elements of ${\bm{z}}(t)$ as the fast states, and elements of ${\bm{x}}(t)$ as the slow states. Other observations, which will prove useful in Sections \ref{subsec:damped} and \ref{subsec:semidamped} are:
	\begin{observation}\label{obs:FastStatePhi}
		The dynamics of $\Phi_{q}$, $\Phi_{d}$, $\Phi_{q}^{(e)}$ and $\Phi_{d}^{(e)}$ are much faster than those of $\Phi_{q_2}$, $\Phi_{d_1}$, $E_{d'}$ and $E_{q'}$.
	\end{observation}
	\begin{observation}\label{obs:FastStateDamp}
		The dynamics of $\Phi_{q_2}$ and $\Phi_{d_1}$ are much faster than those of $E_{d'}$ and $E_{q'}$.
	\end{observation}
	
	\subsection{Classical Model}\label{sec:CLASSm}
	The classical model of a synchronous machine is a second-order model whose formulation is based on the following assumptions \cite{fouad1992power}:
	\begin{inparaenum}[(i)]
		\item the machine can be modeled as a constant magnitude voltage source with a series reactance,
		\item the mechanical rotor angle of the machine can be represented by the angle of the voltage source,
		\item damping can be neglected, and
		\item the machines mechanical power input is constant.
	\end{inparaenum} Thus, the classical model can be obtained from the high-order model by setting $\tau_{q''}=0$, $\tau_{d''}=0$, $\frac{1}{\omega_0}=0$, $\frac{{\omega}^{(s)}(t)}{\omega_0}=1$, $R_{s}=0$, $R_{e}=0$, $X_{q'}=X_{d'}$, $\tau_{q'}=\infty$, $\tau_{d'}=\infty$, $\tau_{m}=\infty$ to give:
	\begin{gather}
	\begin{split}
	\dot{{\delta}}^{(s)}=&\ {\omega}^{(s)}-\omega_0,\\
	M\dot{{\omega}}^{(s)} =&\ T_{m}(0) - \frac{E_0}{X_{d'}^{(e)}}V^{(l)}\sin\left({\delta}^{(s)}-\delta^{(l)}\right) \\& - \tilde{D}_{0}{\omega}^{(s)},
	\end{split}
	\end{gather}
	where $E_0=\sqrt{\left(E_{q'}(0)\right)^2+\left(E_{q'}(0)\right)^2}$ and $X_{d'}^{(e)}\coloneqq X_{d'}+X^{(e)}$ denote constants.
	
	\section{A Library of Second-Order Models}\label{sec:ROm}
	In this section, a library of dynamic models for synchronous machines are developed from the high-order model presented in Section \ref{sec:HOm}. By utilizing the time-scale properties described in Section \ref{sec:observation}, and singular perturbation analysis, the nineteenth-order machine model is reduced to the elemental model, the damped model, and the semi-damped model. Let
	\begin{displaymath}
	R_{s}^{(e)}\coloneqq R_{s}+R^{(e)},\quad X_{q}^{(e)}\coloneqq X_{q}+X^{(e)},\quad X_{d}^{(e)}\coloneqq X_{d}+X^{(e)}.
	\end{displaymath}
	
	\begin{assumption}\label{assum:speed}
		The angular speed of the machine, ${\omega}^{(s)}(t)$, is sufficiently close to the nominal speed of the machine so that $\frac{{\omega}^{(s)}(t)}{\omega_0}=1+\bm{\mathcal{O}}\left(\epsilon\right)$.
	\end{assumption}
	
	\subsection{The Elemental Model}\label{subsec:elem}
	The elemental model is formulated by replacing the differential equations for the fast states with algebraic counterparts called zero-order approximate manifolds. The formulation of these manifolds is presented in Appendix \ref{appendix_elem}.

\begin{figure}[h!t!]
		\centering
		\includegraphics[width=0.95\columnwidth]{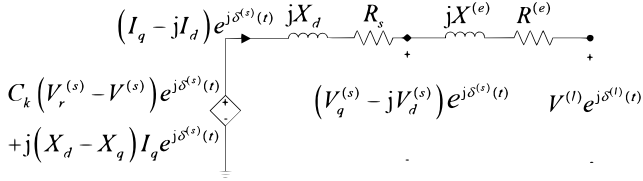}
		\caption{Dynamic circuit of synchronous machine elemental model.}
		\label{fig:EquivCCT_ELEM}
	\end{figure}

	Substituting the zero-order approximate manifolds in \eqref{eqn:zeromanifold} and \eqref{eqn:zeromanifold_1} into \eqref{eqn:fastdamp}--\eqref{eqn:governor_engine}, the elemental model is given by:
	\begin{gather}
	\begin{split}
	\dot{{\delta}}^{(s)}=&\ {\omega}^{(s)}-\omega_0,\\
	M\dot{{\omega}}^{(s)} =&\ P_{r}^{(s)} - {D}_{0}{\omega}^{(s)} - R_{s}^{(e)}\left(I\right)^2 + C_r\left(V^{(l)}\right)^2 \\&-C_{k}C_{r}\left(V_r^{(s)}-V^{(s)}\right)V^{(l)}\cos\left({\delta}^{(s)}-\delta^{(l)}\right)\\& - C_{k}\tilde{C}_{x}\left(V_r^{(s)}-V^{(s)}\right)V^{(l)}\sin\left({\delta}^{(s)}-\delta^{(l)}\right)\\&
	-\frac{C_{x}}{2}\left(V^{(l)}\right)^2\sin2\left({\delta}^{(s)}-\delta^{(l)}\right),
	\end{split}
	\end{gather}
	where $C_r$, $C_{k}$, $\tilde{C}_{x}$ and $C_{x}$ are constants, with \begin{math}
	C_{r} =\frac{R_{s}^{(e)}}{\left(R_{s}^{(e)}\right)^2+X_{q}^{(e)}X_{d}^{(e)}},\;C_{k}=\frac{K_u}{K_f},\;\tilde{C}_{x}=\frac{X_{q}^{(e)}}{\left(R_{s}^{(e)}\right)^2+X_{q}^{(e)}X_{d}^{(e)}},\; C_{x}=\frac{X_{d}-X_{q}}{X_{q}^{(e)}X_{d}^{(e)}},
	\end{math} and \begin{math}
	P_{r}^{(s)}=P_{c}+\bar{D}_{0}\omega_0,\; D_0=\bar{D}_{0}+\tilde{D}_0,\;\text{and}\; I=\sqrt{\left(I_{q}\right)^2+\left(I_{d}\right)^2}.
	\end{math} The dynamic circuit of the elemental model is depicted in Fig. \ref{fig:EquivCCT_ELEM}. For the special case where $R_s$ and $R^{(e)}$ are $\bm{\mathcal{O}}\left(\epsilon\right)$, we set $R_{s}^{(e)}=0$, from where it follows that \begin{math}
	C_{r} =0,\text{ and }\tilde{C}_{x}=\frac{1}{X_{d}^{(e)}}.
	\end{math}

	\subsection{The Damped Model}\label{subsec:damped}
	The damped model is formulated by replacing \eqref{eqn:fastdamp} and \eqref{eqn:slowdamp} with first-order approximate manifolds, and replacing the differential equations for other fast states with zero-order approximate manifolds. By using a first-order approximation for the damper windings manifolds, the effects of damper windings on the machine response are captured by the resulting reduced model. The following simplifying assumption is employed:	
	\begin{assumption}\label{assum:resistance}
		The per-phase line resistance, $R^{(e)}$ is $\bm{\mathcal{O}}\left(\epsilon\right)$.
	\end{assumption}
	\noindent
	Starting with the states observed to have the fastest dynamics, $\Phi_{q}(t)$, $\Phi_{d}(t)$, $\Phi_{q}^{(e)}(t)$ and $\Phi_{d}^{(e)}(t)$, we formulate the zero-order approximate manifolds presented in the first paragraph of Appendix \ref{appendix_damp}.
%	following zero-order approximations by setting $R_s=0$, $R^{(e)}=0$, $\frac{1}{\omega_0}=0$ and $\frac{\omega^{(s)}(t)}{\omega_0}=1$:
%	\begin{equation}
%	\begin{split}
%	\Phi_{q,0}(t) =& - V^{(l)}\sin\left({\delta}^{(s)}(t)-\delta^{(l)}\right),\\
%	\Phi_{d,0}(t) =&\ V^{(l)}\cos\left({\delta}^{(s)}(t)-\delta^{(l)}\right),\\
%	\Phi_{q,0}^{(e)}(t) =&\ V_{d}^{(s)} - V^{(l)}\sin\left({\delta}^{(s)}(t)-\delta^{(l)}\right),\\
%	\Phi_{d,0}^{(e)}(t) =&- V_{q}^{(s)} + V^{(l)}\cos\left({\delta}^{(s)}(t)-\delta^{(l)}\right),
%	\end{split}\label{eqn:PhiZeroMan}
%	\end{equation} from where it follows that: \begin{math}
%	I_{q} = \frac{\left(X_{q'}-X_{q''}\right)}{\left(X_{q'}-X_{k}\right)\left(X_{q''}^{(e)}\right)}\Phi_{q_2}(t)-\frac{\left(X_{q''}-X_{k}\right)}{\left(X_{q'}-X_{k}\right)\left(X_{q''}^{(e)}\right)}E_{d'}(t)+\frac{V^{(l)}\sin\left({\delta}^{(s)}(t)-\delta^{(l)}\right)}{X_{q''}^{(e)}},\end{math} and \begin{math}
%	I_{d} =\ \frac{\left(X_{d'}-X_{d''}\right)}{\left(X_{d'}-X_{k}\right)\left(X_{d''}^{(e)}\right)}\Phi_{d_1}(t)+\frac{\left(X_{d''}-X_{k}\right)}{\left(X_{d'}-X_{k}\right)\left(X_{d''}^{(e)}\right)}E_{q'}(t)-\frac{V^{(l)}\cos\left({\delta}^{(s)}(t)-\delta^{(l)}\right)}{X_{d''}^{(e)}}.
%	\end{math}
	Next, for the subsequent fastest states, $\Phi_{q_2}(t)$ and $\Phi_{d_1}(t)$, which are damper winding states, we derive a first-order approximation of its manifold. Manifolds for $\Phi_{q_2}(t)$ and $\Phi_{d_1}(t)$, can be expressed as power series in $\tau_{q''}$ and $\tau_{d''}$, respectively, to give:
	\begin{equation}
	\begin{split}
	\Phi_{q_2}(t) =&\  \Phi_{q_2,0}(t) + \tau_{q''}\Phi_{q_2,1}(t) + (\tau_{q''})^2\Phi_{q_2,2}(t)+\cdots,\\
	\Phi_{d_1}(t) =&\  \Phi_{d_1,0}(t) + \tau_{d''}\Phi_{d_1,1}(t) + (\tau_{d''})^2\Phi_{d_1,2}(t)+\cdots,
	\end{split}\label{eqn:approxMANx}
	\end{equation}
	where `$0$' subscripts are used to denote a zero-order approximations, and where first-order approximations are given by:
	\begin{equation}
	\begin{split}
	\Phi_{q_2}(t) \approx&\  \Phi_{q_2,0}(t) + \tau_{q''}\Phi_{q_2,1}(t),\\
	\Phi_{d_1}(t) \approx&\  \Phi_{d_1,0}(t) + \tau_{d''}\Phi_{d_1,1}(t).
	\end{split}\label{eqn:approxMAN}
	\end{equation}
	Expressions for $\Phi_{q_2,0}(t)$, $\Phi_{q_2,1}(t)$, $\Phi_{d_1,0}(t)$ and $\Phi_{d_1,1}(t)$ are derived using the following steps:
	\begin{itemize}[$\bullet$]
		\item Substitute \eqref{eqn:approxMANx} into \eqref{eqn:fastdamp} to give:
		\begin{equation}
		\begin{split}
		\tau_{q''} \frac{\mathrm{d}}{\mathrm{d}t}\left(\Phi_{q_2,0}(t) + \tau_{q''}\Phi_{q_2,1}(t)+\cdots\right) = &-\left(\Phi_{q_2,0}(t)\right.\\\left. + \tau_{q''}\Phi_{q_2,1}(t)+\cdots\right) - \left(X_{q'}-X_{k}\right)&I_q - E_{d'}(t),\\
		\tau_{d''} \frac{\mathrm{d}}{\mathrm{d}t}\left(\Phi_{d_1,0}(t) + \tau_{d''}\Phi_{d_1,1}(t)+\cdots\right) = &-\left(\Phi_{d_1,0}(t)\right.\\\left. + \tau_{d''}\Phi_{d_1,1}(t)+\cdots\right) - \left(X_{d'}-X_{k}\right)&I_d +  E_{q'}(t).
		\end{split}\label{eqn:manCALC}
		\end{equation}
		\item Using the zero-order approximations in \eqref{eqn:PhiZeroMan}, substitute expressions for $I_q$ and $I_d$ into \eqref{eqn:manCALC} and equate the $\left(\tau_{q''}\right)^0$ and $\left(\tau_{d''}\right)^0$ terms to give: \begin{equation*}
		\begin{split}
		\Phi_{q_2,0}(t) =&-\frac{X_{k}^{(e)}}{X_{q'}^{(e)}}E_{d'}(t)-\frac{X_{q'}-X_{k}}{X_{q'}^{(e)}}V^{(l)}\sin\left({\delta}^{(s)}(t)-\delta^{(l)}\right),\\
		\Phi_{d_1,0}(t) =&\ \frac{X_{k}^{(e)}}{X_{d'}^{(e)}}E_{q'}(t)+\frac{X_{d'}-X_{k}}{X_{d'}^{(e)}}V^{(l)}\cos\left({\delta}^{(s)}(t)-\delta^{(l)}\right),
		\end{split}
		\end{equation*}
		where \begin{math} X_{k}^{(e)}\coloneqq X_{k}+X^{(e)}.
		\end{math}
		\item Also equate the $\left(\tau_{q''}\right)^1$ and $\left(\tau_{d''}\right)^1$ terms to give:
		\begin{align*}
		\Phi_{q_2,1}(t) =&-\frac{X_{q''}^{(e)}X_{k}^{(e)}}{\tau_{q'}\left(X_{q'}^{(e)}\right)^3}\left(X_{q}^{(e)}E_{d'}(t)-\left(X_{q}-X_{q'}\right)V^{(l)}\right.\\&\left.\cdot\sin\left({\delta}^{(s)}(t)-\delta^{(l)}\right)\right)+\frac{X_{q''}^{(e)}\left(X_{q'}-X_{k}\right)}{\left(X_{q'}^{(e)}\right)^2}\dot{V}_{d}^{(l)},\\
		\Phi_{d_1,1}(t) =&\ \frac{X_{d''}^{(e)}X^{(e)}_{k}}{\tau_{d'}\left(X_{d'}^{(e)}\right)^3}\left(X_{d}^{(e)}E_{q'}(t)-\left(X_{d}-X_{d'}\right)V^{(l)}\right.\\&\left.\cdot\cos\left({\delta}^{(s)}(t)-\delta^{(l)}\right)\right)-\frac{X_{d''}^{(e)}\left(X_{d'}-X_{k}\right)}{\left(X_{d'}^{(e)}\right)^2}\dot{V}_{q}^{(l)}\\&-\frac{X_{d''}^{(e)}X_{k}^{(e)}}{\tau_{d'}\left(X_{d'}^{(e)}\right)^2}E_f(t),
		\end{align*}
	\end{itemize}
	where \begin{math}\dot{V}_{d}^{(l)}=V^{(l)}\cos\left({\delta}^{(s)}(t)-\delta^{(l)}\right)\left(\dot{\delta}^{(s)}(t)-\dot{\delta}^{(l)}\right)+\dot{V}^{(l)}\sin\left({\delta}^{(s)}(t)-\delta^{(l)}\right),\; \dot{V}_{q}^{(l)}=\dot{V}^{(l)}\cos\left({\delta}^{(s)}(t)-\delta^{(l)}\right)-V^{(l)}\sin\left({\delta}^{(s)}(t)-\delta^{(l)}\right)\left(\dot{\delta}^{(s)}(t)-\dot{\delta}^{(l)}\right), \text{ and } X_{q'}^{(e)}\coloneqq X_{q'}+X^{(e)}.\end{math} Next, for the damper winding state observed to have the slower dynamics, $E_{d'}(t)$, we derive a first-order approximation of its manifold. A manifold for $E_{d'}(t)$ can be expressed as a power series in $\tau_{q'}$ to give:
	\begin{equation}
	E_{d'}(t) =\  E_{d',0}(t) + \tau_{q'}E_{d',1}(t) + (\tau_{q'})^2E_{d',2}(t)+\cdots,\label{eqn:approxMANx2}
	\end{equation}
	from where it follows that a first-order approximation is given by:
	\begin{equation}
	\begin{split}
	E_{d'}(t) \approx&\  E_{d',0}(t) + \tau_{q'}E_{d',1}(t).
	\end{split}\label{eqn:approxMAN2}
	\end{equation}
	Expressions for $E_{d',0}(t)$ and $E_{d',1}(t)$ can be derived using the following steps:
	\begin{itemize}[$\bullet$]
		\item Substitute \eqref{eqn:approxMAN} and \eqref{eqn:approxMANx2} into \eqref{eqn:slowdamp} to give:
		\begin{equation}
		\begin{split}
		\tau_{q'}&\frac{\mathrm{d}}{\mathrm{d}t}\left(E_{d',0}(t) + \tau_{q'}E_{d',1}(t)+\cdots\right)=\\& -\left(E_{d',0}(t)+\tau_{q'}E_{d',1}(t)+\cdots\right)\\&+\left(X_q-X_{q'}\right)\left(I_q-\frac{X_{q'}\left(X_{q'}-X_{q''}\right)}{X_{q''}\left(X_{q'}-X_{k}\right)^2}\tau_{q''}\Phi_{q_2,1}(t)\right).
		\end{split}\label{eqn:manCALC2}
		\end{equation}
		\item Using the zero-order approximations in \eqref{eqn:PhiZeroMan}, substitute the expressions for $I_q$ and $\Phi_{q_2,1}(t)$, into \eqref{eqn:manCALC2}, and equate the $\left(\tau_{q'}\right)^0$ terms to give:
		\begin{equation*}
		\begin{split}
		E_{d',0}(t) =&\ \frac{X_{q}-X_{q'}}{X_{q}^{(e)}}V^{(l)}\sin\left({\delta}^{(s)}(t)-\delta^{(l)}\right)-\frac{N_{q}}{D_{q}}\dot{V}_{d}^{(l)},
		\end{split}
		\end{equation*}
		where \begin{math}
		N_{q} = \tau_{q'}\tau_{q''}X_{q'}^{(e)}X_{k}^{(e)}\left(X_{q}-X_{q'}\right)\left(X_{q'}-X_{q''}\right)\\\cdot\left(X_{q'}-X_{k}\right),\;	D_{q} = \tau_{q'}X_{q}^{(e)}\left(X_{q'}^{(e)}\right)^2\left(X_{q'}-X_{k}\right)^2-\tau_{q''}X_{q}^{(e)}\left(X_{k}^{(e)}\right)^2\left(X_{q}-X_{q'}\right)\left(X_{q'}-X_{q''}\right).
		\end{math}
		\item Also equate the $\left(\tau_{q'}\right)^1$ terms to give:
		\begin{equation*}
		\begin{split}
		E_{d',1}(t) =&
		-\frac{N_{q'}}{\tilde{D}_{q}}\dot{V}_{d}^{(l)} + \bm{\mathcal{O}}\left(\tau_{q'}\right),
		\end{split}
		\end{equation*}
		where \begin{math}
		N_{q'} = \tau_{q'}\left(X_{q'}^{(e)}\right)^3\left(X_{q}-X_{q'}\right)\left(X_{q'}-X_{k}\right)^2\text{ and }\\
		\tilde{D}_{q} = X_{q}^{(e)}D_{q}.
		\end{math}
	\end{itemize}
	Finally, for other states observed to have fast dynamics, i.e., $E_{q'}$, $E_{f}$, $U_{f}$, $\bar{U}_{f}$, $T_{m}$, $P_{u}$, $P_{a_2}$, $P_{b_2}$, $P_{a_1}$, and $P_{b_1}$, zero-order manifolds are derived as described in the second paragraph of Appendix \ref{appendix_damp}.
%	following zero-order approximations  by setting $\tau_{d'}$, and all $\bm{\mathcal{O}}\left(\epsilon\right)$ parameters except $\tau_{q''}$ and $\tau_{q'}$, to zero:
%	\begin{equation}
%	\begin{split}
%	E_{f,0}(t) =&\ \frac{K_u\left(V_r^{(s)}-V^{(s)}\right)}{K_f},\\
%	E_{q',0}(t) =&\ \frac{X_{d'}^{(e)}}{X_{d}^{(e)}}E_{f,0}(t)-\frac{N_{d}}{D_{d}}\dot{V}_q^{(l)}\\&+\frac{X_d-X_{d'}}{X_d^{(e)}}V^{(l)}\cos\left({\delta}^{(s)}(t)-\delta^{(l)}\right),\\
%	U_{f,0}(t) =&\ K_fE_{f,0}(t),\\
%	\bar{U}_{f,0}(t) =&\ \frac{\bar{K}_u}{\bar{\tau}_u}E_{f,0}(t),\\
%	P_{u,0}(t) =&\ P_{c} - \bar{D}_{0}\left({{\omega}}^{(s)}(t)-\omega_0\right),\\
%	T_{m,0}(t) =&\ P_{u,0}(t),
%	\end{split}\label{eqn:zeromanifold2}
%	\end{equation}
%	also, $P_{a_1,0} = P_{a_2,0} = P_{b_1,0} = P_{b_2,0} = 0$, where \begin{math}
%	N_{d} = \tau_{d'}\tau_{d''}X_{d'}^{(e)}X_{k}^{(e)}\left(X_{d}-X_{d'}\right)\left(X_{d'}-X_{d''}\right)\left(X_{d'}-X_{k}\right),\; \text{and}\\D_{d} = \tau_{d'}X_{d}^{(e)}\left(X_{d'}^{(e)}\right)^2\left(X_{d'}-X_{k}\right)^2-\tau_{d''}X_{d}^{(e)}\left(X_{k}^{(e)}\right)^2\\\cdot\left(X_{d}-X_{d'}\right)\left(X_{d'}-X_{d''}\right).
%	\end{math}

\begin{figure}[h!t!]
		\centering
		\includegraphics[width=0.95\columnwidth]{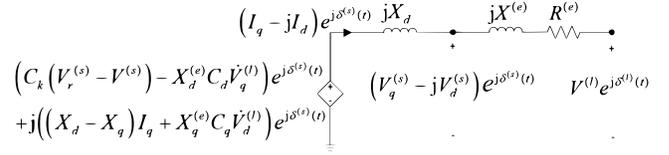}
		\caption{Dynamic circuit of synchronous machine damped model.}
		\label{fig:EquivCCT_DAMP}
	\end{figure}

	Substituting the first-order approximate manifolds in \eqref{eqn:approxMAN}, \eqref{eqn:approxMAN2}, and the zero-order approximate manifolds in \eqref{eqn:zeromanifold}, \eqref{eqn:PhiZeroMan}, and \eqref{eqn:zeromanifold2} into \eqref{eqn:fastdamp}--\eqref{eqn:governor_engine}, and setting $\bm{\mathcal{O}}\left((\tau_{q'})^2\right)$ terms to zero, the damped model for a non-salient pole machine is given by:
	\begin{gather}
	\begin{split}
	\dot{{\delta}}^{(s)}=&\ {\omega}^{(s)}-\omega_0,\\
	M\dot{\omega}^{(s)} =&\ P_{r}^{(s)} - {D}_{0}{\omega}^{(s)}-\frac{C_{x}}{2}\left(V^{(l)}\right)^2\sin 2\left({\delta}^{(s)}-\delta^{(l)}\right)
	\\&-\frac{C_{k}}{X_{d}^{(e)}}\left(V_r^{(s)}-V^{(s)}\right)V^{(l)}\sin\left({\delta}^{(s)}-\delta^{(l)}\right)\\& - C_{q}\left(V^{(l)}\right)^2\cos^2\left({\delta}^{(s)}-\delta^{(l)}\right)\left(\dot{\delta}^{(s)}-\dot{\delta}^{(l)}\right)\\& -
	C_{d}\left(V^{(l)}\right)^2\sin^2\left({\delta}^{(s)}-\delta^{(l)}\right)\left(\dot{\delta}^{(s)}-\dot{\delta}^{(l)}\right)
	\\& - \frac{\left(C_{q}-C_{d}\right)}{2}\dot{V}^{(l)}{V}^{(l)}\sin 2\left({\delta}^{(s)}-\delta^{(l)}\right),
	\end{split}
	\end{gather}
	where \begin{math}
	C_{k},\;C_{x},\;C_{q},\;\text{and}\;C_{d}\end{math} are constants, \begin{math}
	C_{k}=\frac{K_u}{K_f},\;C_{x}=\frac{\left(X_{d}-X_{q}\right)}{X_{q}^{(e)}X_{d}^{(e)}},
	\end{math} \begin{math}
	C_{q}=C_{q''}+\left(C_{q'}+\tilde{C}_{q''}\right)^2\tilde{C}_{q},
	\end{math} with \begin{math}
	C_{q''}=\frac{\tau_{q''}\left(X_{q'}-X_{q''}\right)}{\left(X_{q'}^{(e)}\right)^2},\;\tilde{C}_{q}=\frac{\left(X_{q}-X_{q'}\right)}{\tilde{D}_{q}},\;C_{q'}=\tau_{q'}X_{q'}^{(e)}\left(X_{q'}-X_{k}\right),\;\tilde{C}_{q''}=\frac{\tau_{q''}X_{q}^{(e)}X_{k}^{(e)}\left(X_{q'}-X_{q''}\right)}{X_{q'}^{(e)}},
	\end{math} and \begin{math}C_{d}=C_{d''}+\left(C_{d'}+\tilde{C}_{d''}\right)\tilde{C}_{d''}\tilde{C}_{d},
	\end{math} with \begin{math}	C_{d''}=\frac{\tau_{d''}\left(X_{d'}-X_{d''}\right)}{\left(X_{d'}^{(e)}\right)^2},\;\tilde{C}_{d}=\frac{\left(X_{d}-X_{d'}\right)}{\tilde{D}_{d}},\;C_{d'}=\tau_{d'}X_{d'}^{(e)}\left(X_{d'}-X_{k}\right),\;\tilde{C}_{d''}=\frac{\tau_{d''}X_{d}^{(e)}X_{k}^{(e)}\left(X_{d'}-X_{d''}\right)}{X_{d'}^{(e)}}.
	\end{math}The dynamic circuit of the damped model is depicted in Fig. \ref{fig:EquivCCT_DAMP}, with \begin{math}\dot{V}_{d}^{(l)}=V^{(l)}\cos\left({\delta}^{(s)}(t)-\delta^{(l)}\right)\left(\dot{\delta}^{(s)}(t)-\dot{\delta}^{(l)}\right)+\dot{V}^{(l)}\sin\left({\delta}^{(s)}(t)-\delta^{(l)}\right),\;\dot{V}_{q}^{(l)}=\dot{V}^{(l)}\cos\left({\delta}^{(s)}(t)-\delta^{(l)}\right)-V^{(l)}\sin\left({\delta}^{(s)}(t)-\delta^{(l)}\right)\left(\dot{\delta}^{(s)}(t)-\dot{\delta}^{(l)}\right).\end{math} Note that for salient pole machines, $\tilde{C}_{q}=0$, whereas for round-rotor machines, $C_{x}=0$.
	
	\subsection{The Semi-Damped Model}\label{subsec:semidamped}
	The semi-damped model is developed by replacing \eqref{eqn:slowdamp} with a first-order approximate manifold, and replacing the differential equations for other fast states with zero-order approximate manifolds. In salient-pole machines, only one damper winding is aligned with the $q$-axis, and the damper winding represented by \eqref{eqn:slowdamp} is typically excluded \cite{KrauseWasynczuk2013}. Due to this reason, the semi-damped model is only applicable to round-rotor machines.
	
	Starting with fastest states $\Phi_{q}(t)$, $\Phi_{d}(t)$, $\Phi_{q}^{(e)}(t)$, $\Phi_{d}^{(e)}(t)$, $\Phi_{q_2}(t)$ and $\Phi_{d_1}(t)$, we develop the zero-order approximate manifolds presented in the first paragraph of Appendix \ref{appendix_semidamp}.
%	following zero-order approximations by setting $R_s=0$, $R^{(e)}=0$, $\frac{1}{\omega_0}=0$, $\frac{\omega^{(s)}(t)}{\omega_0}=1$, $\tau_{q''}=0$ and $\tau_{d''}=0$:
%	\begin{equation}
%	\begin{split}
%	\Phi_{q,0}(t) =& - V^{(l)}\sin\left({\delta}^{(s)}(t)-\delta^{(l)}\right),\\
%	\Phi_{d,0}(t) =&\ V^{(l)}\cos\left({\delta}^{(s)}(t)-\delta^{(l)}\right),
%	\end{split}\label{eqn:PhiZeroMan1}
%	\end{equation}
%	\begin{equation}
%	\begin{split}
%	\Phi_{q,0}^{(e)}(t) =&\ V_{d}^{(s)} - V^{(l)}\sin\left({\delta}^{(s)}(t)-\delta^{(l)}\right),\\
%	\Phi_{d,0}^{(e)}(t) =&- V_{q}^{(s)} + V^{(l)}\cos\left({\delta}^{(s)}(t)-\delta^{(l)}\right),\\
%	\Phi_{q_2,0}(t) =&-\left(X_{q'}-X_{k}\right)I_{q}-E_{d'}(t),\\
%	\Phi_{d_1,0}(t) =& -\left(X_{d'}-X_{k}\right)I_{d} + E_{q'}(t),
%	\end{split}\label{eqn:PhiZeroMan2}
%	\end{equation}
%	from where it follows that: \begin{math}
%	I_{q} =-\frac{1}{X_{q'}^{(e)}}E_{d'}(t)+\frac{V^{(l)}\sin\left({\delta}^{(s)}(t)-\delta^{(l)}\right)}{X_{q'}^{(e)}},\;
%	I_{d} = \frac{1}{X_{d'}^{(e)}}E_{q'}(t)-\frac{V^{(l)}\cos\left({\delta}^{(s)}(t)-\delta^{(l)}\right)}{X_{d'}^{(e)}}.
%	\end{math}	
	Next, we derive a first-order approximate manifold for $E_{d'}(t)$ having the form \begin{align}
	E_{d'}(t) \approx\  E_{d',0}(t) + \tau_{q'}E_{d',1}(t).\label{eqn:1storder_semi}
	\end{align} Substituting the expression for $\Phi_{q_2,0}(t)$ in \eqref{eqn:PhiZeroMan1}, and the power series expansion in \eqref{eqn:approxMANx2} into \eqref{eqn:slowdamp}, it follows that:
	\begin{equation}
	\begin{split}
	\tau_{q'}\frac{\mathrm{d}}{\mathrm{d}t}&\left(E_{d',0}(t) + \tau_{q'}E_{d',1}(t)+\cdots\right)=\\& -\left(E_{d',0}(t)+\tau_{q'}E_{d',1}(t)+\cdots\right)\frac{X_{q}^{(e)}}{X_{q'}^{(e)}}\\&+\frac{X_q-X_{q'}}{X_{q'}^{(e)}}V^{(l)}\sin\left({\delta}^{(s)}(t)-\delta^{(l)}\right).
	\end{split}\label{eqn:manCALC3}
	\end{equation}
	Equating the $\left(\tau_{q'}\right)^0$ terms in \eqref{eqn:manCALC3}, we have that:
	\begin{align}
	E_{d',0}(t) =&\ \frac{X_{q}-X_{q'}}{X_{q}^{(e)}}V^{(l)}\sin\left({\delta}^{(s)}(t)-\delta^{(l)}\right),
	\end{align}
	and equating the $\left(\tau_{q'}\right)^1$ terms, we have that:\begin{align}
	E_{d',1}(t) =&
	-\frac{X_{q'}^{(e)}\left(X_{q}-X_{q'}\right)}{\left(X_{q}^{(e)}\right)^2}\dot{V}_{d}^{(l)},
	\end{align}where \begin{math}\dot{V}_{d}^{(l)}=V^{(l)}\cos\left({\delta}^{(s)}(t)-\delta^{(l)}\right)\left(\dot{\delta}^{(s)}(t)-\dot{\delta}^{(l)}\right)+\dot{V}^{(l)}\sin\left({\delta}^{(s)}(t)-\delta^{(l)}\right).\end{math}	
	Finally, for other states observed to have fast dynamics, i.e., $E_{q'}$, $E_{f}$, $U_{f}$, $\bar{U}_{f}$, $T_{m}$, $P_{u}$, $P_{a_2}$, $P_{b_2}$, $P_{a_1}$ and $P_{b_1}$, zero-order approximate manifolds are derived as described in the second paragraph of Appendix \ref{appendix_semidamp}.
%	following zero-order approximations, derived by setting $\tau_{d''}$, $\tau_{d'}$, and all $\bm{\mathcal{O}}\left(\epsilon\right)$ parameters except $\tau_{q'}$, to zero are used:
%	\begin{equation}
%	\begin{split}
%	E_{f,0}(t) =&\ \frac{K_u\left(V_r^{(s)}-V^{(s)}\right)}{K_f},\\
%	E_{q',0}(t) =&\ \frac{X_{d'}^{(e)}}{X_{d}^{(e)}}E_{f,0}(t)+\frac{X_d-X_{d'}}{X_d^{(e)}}V^{(l)}\cos\left({\delta}^{(s)}(t)-\delta^{(l)}\right),\\
%	U_{f,0}(t) =&\ K_fE_{f,0}(t),\\
%	\bar{U}_{f,0}(t) =&\ \frac{\bar{K}_u}{\bar{\tau}_u}E_{f,0}(t),\\
%	P_{u,0}(t) =&\ P_{c} - \bar{D}_{0}\left({{\omega}}^{(s)}(t)-\omega_0\right),\\
%	T_{m,0}(t) =&\ P_{u,0}(t),
%	\end{split}\label{eqn:zeromanifold3}
%	\end{equation}
%	with, $P_{a_1,0} = P_{a_2,0} = P_{b_1,0} = P_{b_2,0} = 0$.
	
\begin{figure}[h!t!]
		\centering
		\includegraphics[width=0.95\columnwidth]{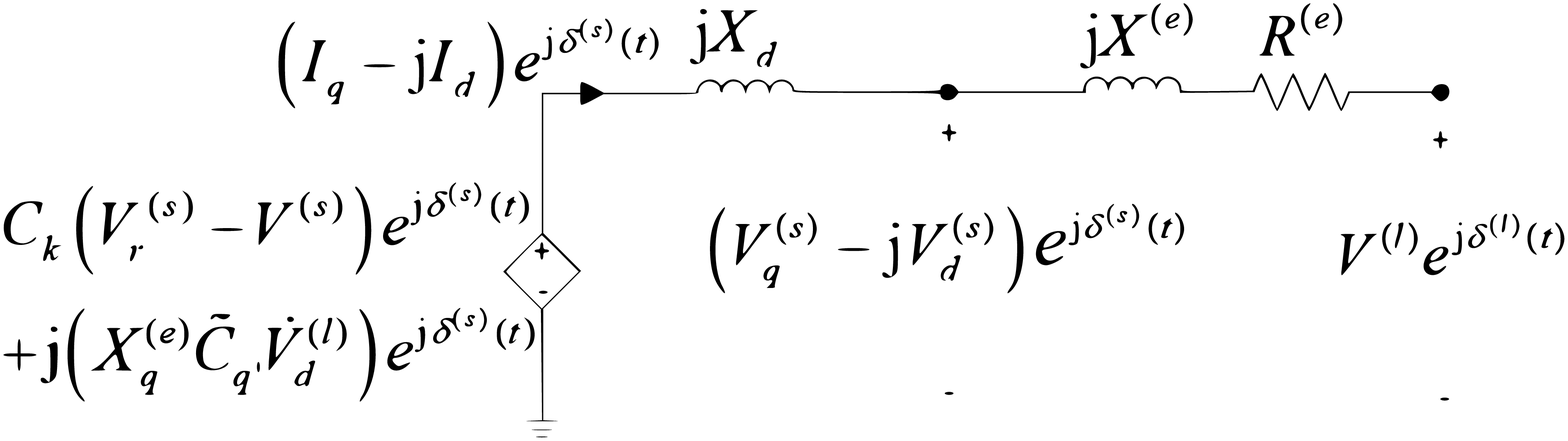}
		\caption{Dynamic circuit of synchronous machine semi-damped model.}
		\label{fig:EquivCCT_SEMIDAMP}
	\end{figure}

	Substituting the zero-order approximate manifolds in \eqref{eqn:zeromanifold}, \eqref{eqn:PhiZeroMan1}, \eqref{eqn:zeromanifold3}, and the first-order approximate manifold in \eqref{eqn:1storder_semi} into \eqref{eqn:fastdamp}--\eqref{eqn:governor_engine}, the semi-damped model is given by:
	\begin{gather}
	\begin{split}
	\dot{{\delta}}^{(s)}=&\ {\omega}^{(s)}-\omega_0,\\
	M\dot{\omega}^{(s)} =&\ P_{r}^{(s)} - {D}_{0}{\omega}^{(s)} - \frac{\tilde{C}_{q'}}{2}\dot{V}^{(l)}{V}^{(l)}\sin 2\left({\delta}^{(s)}-\delta^{(l)}\right)\\& - \tilde{C}_{q'}\left(V^{(l)}\right)^2\cos^2\left({\delta}^{(s)}-\delta^{(l)}\right)\left(\dot{\delta}^{(s)}-\dot{\delta}^{(l)}\right)\\&-\frac{C_{k}}{X_{d}^{(e)}}\left(V_r^{(s)}-V^{(s)}\right)V^{(l)}\sin\left({\delta}^{(s)}-\delta^{(l)}\right),
	\end{split}
	\end{gather}
	where \begin{math}
	C_{k}\;\text{and}\;\tilde{C}_{q'}\end{math} are constants, with \begin{math}
	C_{k}=\frac{K_u}{K_f},\;\text{and}\;	\tilde{C}_{q'}=\frac{\tau_{q'}\left(X_{q}-X_{q'}\right)}{\left(X_{q}^{(e)}\right)^2}.
	\end{math} The dynamic circuit of the semi-damped model is depicted in Fig. \ref{fig:EquivCCT_SEMIDAMP}, with \begin{math}\dot{V}_{d}^{(l)}=V^{(l)}\cos\left({\delta}^{(s)}(t)-\delta^{(l)}\right)\left(\dot{\delta}^{(s)}(t)-\dot{\delta}^{(l)}\right)+\dot{V}^{(l)}\sin\left({\delta}^{(s)}(t)-\delta^{(l)}\right).\end{math}

	\section{Numerical Validation}\label{sec:simRes}
\begin{figure}[h!t!]
		\centering
		\includegraphics[width=0.825\columnwidth]{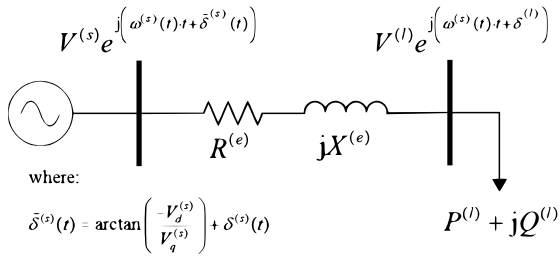}
		\caption{One line diagram of a power system with a synchronous machine connected to a constant power load through a short transmission line.}
		\label{fig:two_bus}
	\end{figure}

	In this section, simulation results comparing the high-order model, the classical model, the elemental model, the semi-damped model, and the damped model, of a round-rotor synchronous machine, are presented. We consider a two-bus power system with a synchronous machine connected to a constant power load through a short electrical transmission line. See Fig. \ref{fig:two_bus} for a one-line diagram, and Table \ref{tab:MODparam} for the system parameters.
	
	\begin{figure*}[h!t!]
		\centering
		\includegraphics[width=0.67\columnwidth]{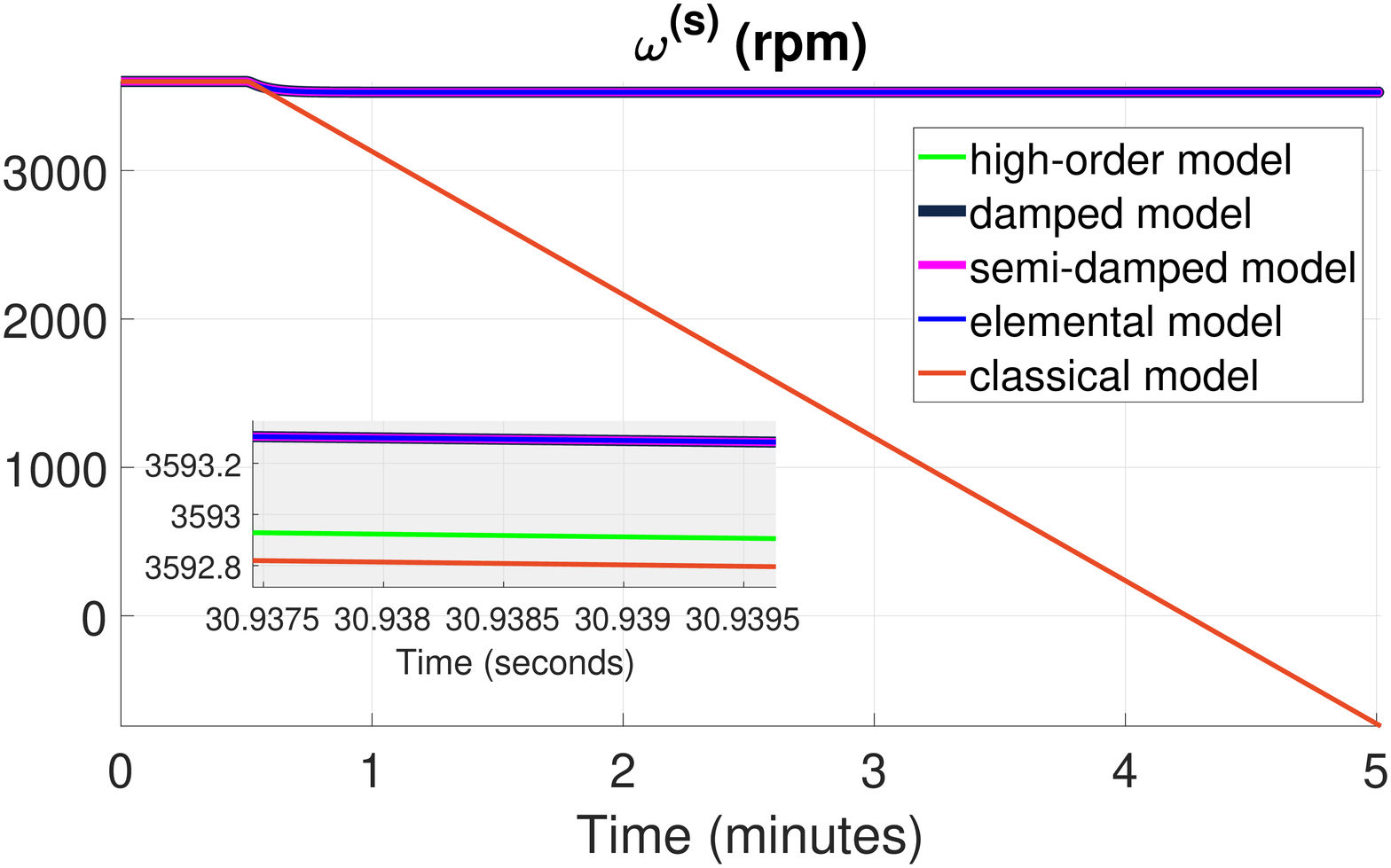}
		\includegraphics[width=0.67\columnwidth]{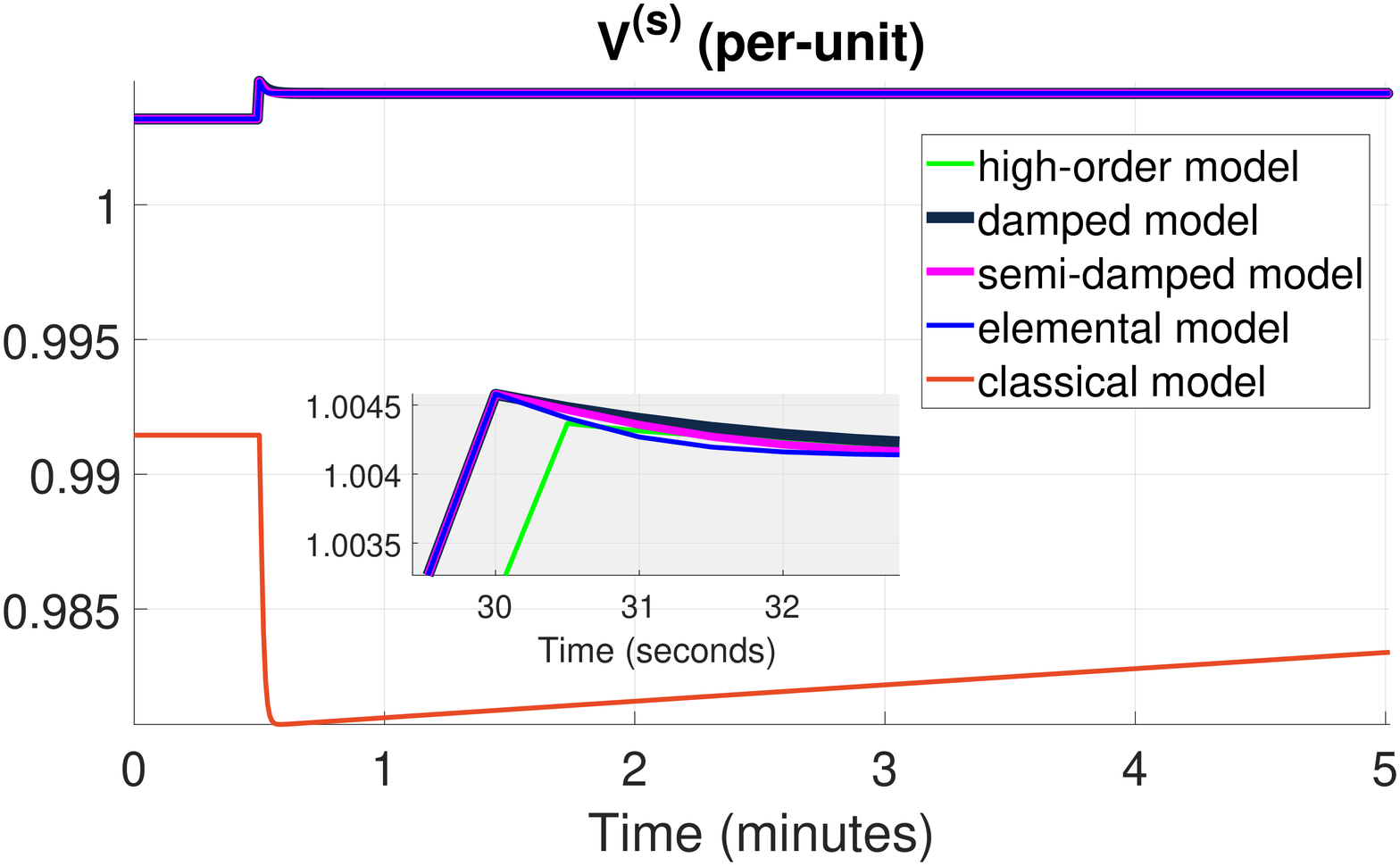}
		\includegraphics[width=0.67\columnwidth]{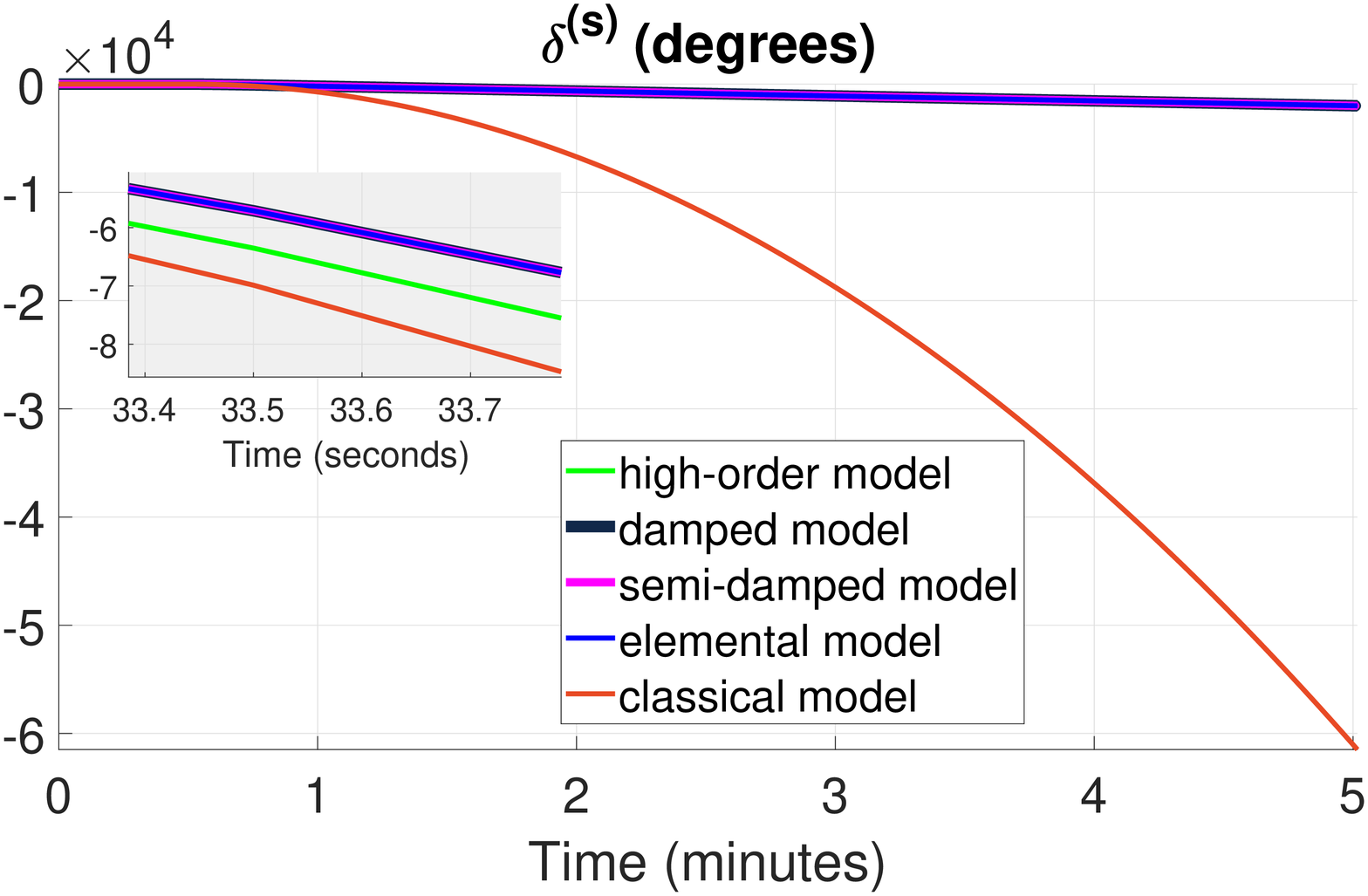}
		\caption{Case 1 numerical results: machine angular frequency, voltage magnitude and phase.}
		\label{fig:Case_1}
	\end{figure*}
	\begin{figure*}[h!t!]
		\centering
		\includegraphics[width=0.67\columnwidth]{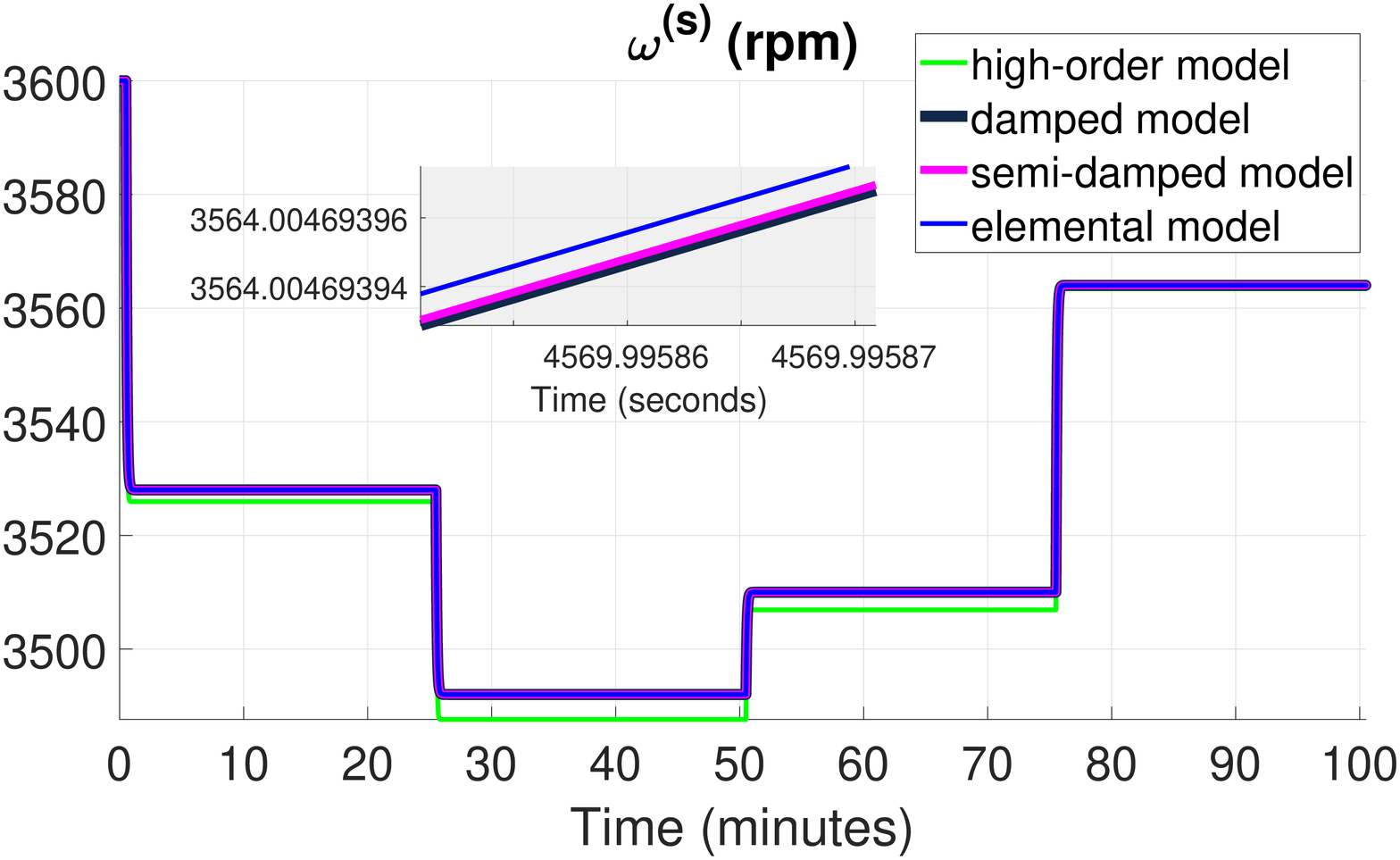}
		\includegraphics[width=0.67\columnwidth]{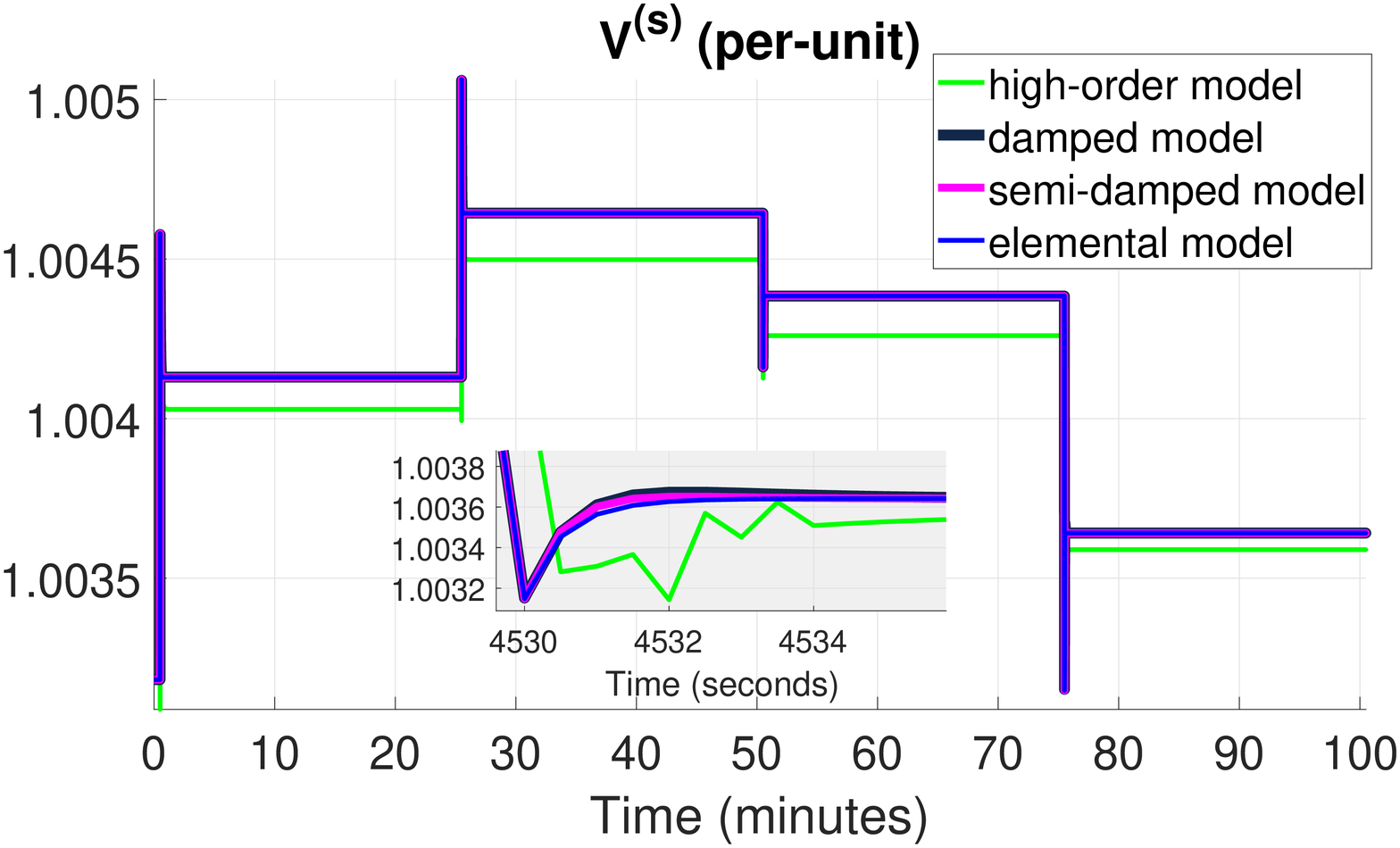}
		\includegraphics[width=0.67\columnwidth]{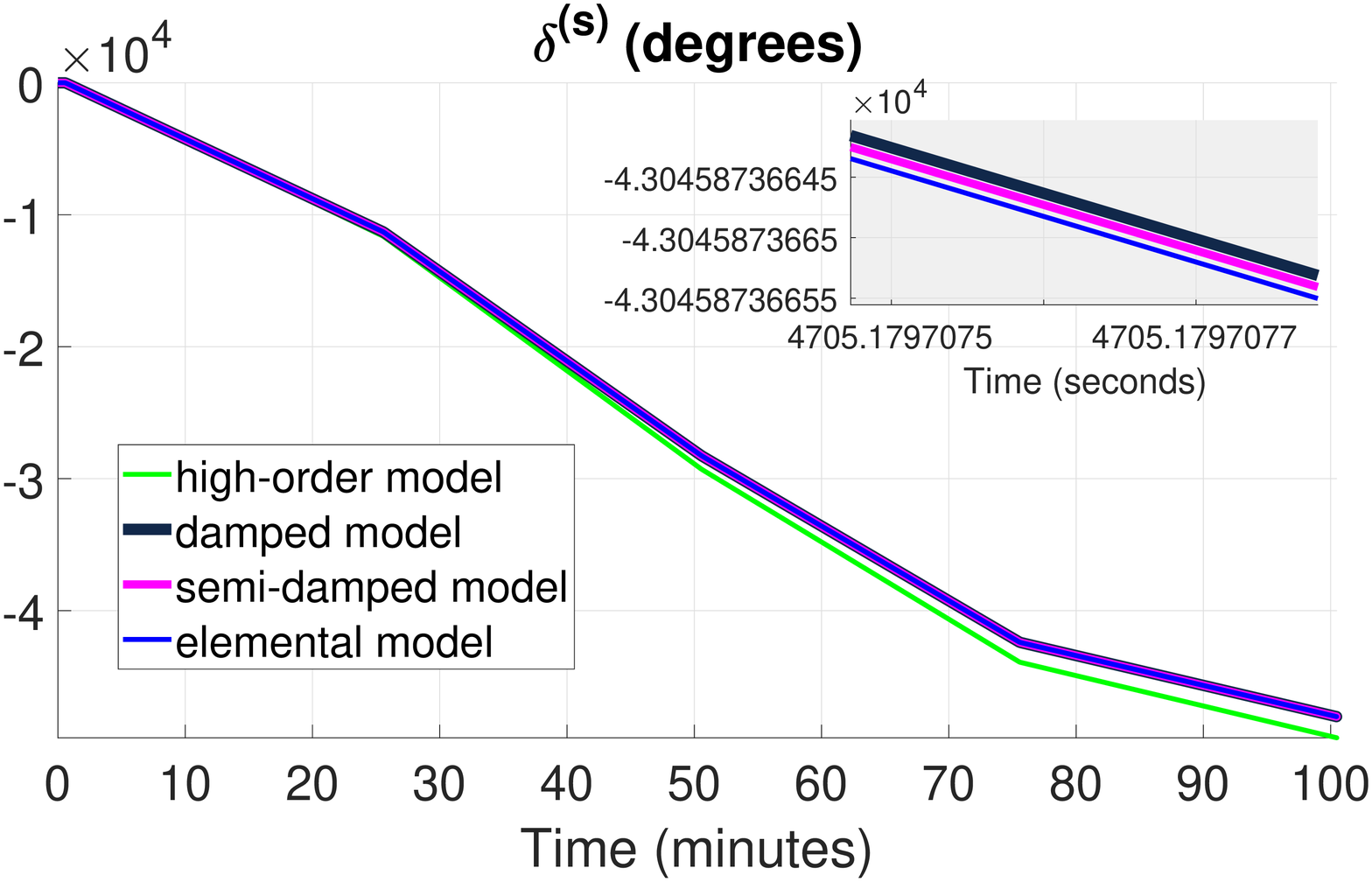}
		\caption{Case 2 numerical results: machine angular frequency, voltage magnitude and phase.}
		\label{fig:Case_2}
	\end{figure*}
	\subsection{Case 1}
	This case is used to highlight the high-fidelity of the second-order models in comparison to the classical model, and we consider the system response to an increase in real power demand by the load. A stable equilibrium point of the high-order model is chosen as the common initial condition for all the models. The real power demand by the load is increased from 0.05 [pu] to 0.25 [pu] at time $t=30$ [s], and the reference voltage magnitude $V_r^{(s)}$ is changed at time $t=30$ [s] to keep the bus voltage magnitude at $V^{(l)}=1$ [pu]. Numerical results are depicted in Fig. \ref{fig:Case_1}, and they show that the elemental model, the semi-damped model and the damped model have an overall better accuracy than the classical model, and that after one second, the error of the classical model response increases exponentially.
	\subsection{Case 2}
	This case is used to compare the fidelity of the elemental model, the semi-damped model and the damped model. The machine whose parameters are described in Table \ref{tab:MODparam} is employed. The real power demand by the load is increased from 0.05 [pu] to 0.25 [pu] at time $t=30$ [s], from 0.25 [pu] to 0.35 [pu] at time $t=1530$ [s], from 0.35 [pu] to 0.3 [pu] at time $t=3030$ [s], and from 0.3 [pu] to 0.15 [pu] at time $t=4530$ [s]. For each load change, the reference voltage magnitude $V_r^{(s)}$ is changed to keep the bus voltage magnitude at $V^{(l)}=1$ [pu]. The root mean square errors of the models, relative to the high-order model, are outlined in Table \ref{tab:RMSE}, and numerical results are presented in Figs. \ref{fig:Case_2}.
	\begin{center}
		\begin{table}[ht]
			\centering
			\caption{Root Mean Square Error (RMSE)}
			\begin{tabular}{|l|c|c|c|}
				\hline
				& {$\omega^{(s)}$} & {$V^{(s)}$} & {$\delta^{(s)}$} \\ \hline
				{damped model} & $2.9093$[rpm] &$0.0042$ [pu] &$1065.6$ [deg] \\ {semi-damped model} & $2.9093$[rpm] &$0.0042$ [pu] &$1065.6$ [deg] \\ {elemental model}& $80.753$[rpm] &$0.0042$ [pu] &$1065.6$ [deg] \\
				\hline
			\end{tabular} \label{tab:RMSE}
		\end{table}
	\end{center}
	The RMSE results show that although the elemental model, the semi-damped model and the damped model match in accuracy for machine voltage magnitude response, the damped model and the semi-damped model have a higher accuracy for machine angular frequency response.
	
	\begin{center}
		\begin{table}[ht]
			\centering
			\caption{System parameters for a salient pole synchronous machine}
			\begin{tabular}{|l|c|c|}
				\hline
				& \textit{parameter} & \textit{value} \\ \hline
				\multirow{6}{*}{Damper windings} & $\tau_{q''}$ &0.9453 [s] \\ &$\tau_{d''}$ &0.042 [s] \\ & $\tau_{q'}$ &3.6123 [s] \\ &$X_{q''}$ &0.2388 [pu] \\ &$X_{q'}$ &0.7299 [pu] \\ &$X_{q}$ &1.7997 [pu] \\ &$X_{k}$ &0.19 [pu] \\
				\hline
				\multirow{4}{*}{Stator windings} &$\omega_0$ &376.99 [rad/s] \\ &$R_s$ &0.003 [pu] \\ &$X_{d''}$ &0.24 [pu] \\ &$X_{d'}$ &0.32 [pu] \\
				\hline
				\multirow{6}{*}{IEEE DC1A exciter} & $\tau_{d'}$ &5.0141 [s] \\ &$\tau_{f}$ &$1\times10^{-8}$ [s] \\ &$\tau_{u}$ &0.002 [s] \\ &$\bar{\tau}_{u}$ &$1\times10^{-12}$ [s] \\ &$X_{d}$ &1.7997 [pu]\\ &$K_{f}$ &1 [pu] \\ &$K_{u}$ &200 \\ &$\bar{K}_{u}$ &0 [s] \\
				\hline
				\multirow{11}{*}{DEGOV1 speed governor} & $\tau_{1}$ &$1\times10^{-4}$ [s] \\ &$\tau_{2}$ &0 [s] \\ &$\tau_{3}$ &0.5001 [s] \\ &$\tau_{4}$ &$25\times10^{-3}$ [s] \\ &$\tau_{5}$ &$9\times10^{-4}$ [s] \\ &$\tau_{6}$ &$5.74\times10^{-3}$ [s] \\ &$\tau_{m}$ &$24\times10^{-3}$ [s] \\ &$\kappa$ &10 \\ &$P_r^{(s)}$ &0 [pu] \\ &$M$ &0.1188 [s$^2$] \\ &$\tilde{D}_{0}$ &$2.5825\times10^{-7}$ [s/rad] \\ &$\bar{D}_{0}$ &0.0531 [s/rad] \\\hline
				\multirow{2}{*}{Transmission line} &$R^{(e)}$ &0.004 [pu] \\ &$X^{(e)}$ &0.0595 [pu] \\
				\hline
			\end{tabular} \label{tab:MODparam}
		\end{table}
	\end{center}

	\section{Concluding Remarks}\label{sec:conclusion}
	In this paper, we introduced a library of second-order synchronous machine models, comprising of the elemental model, the damped model, and the semi-damped model. We also showed how these models, and the so-called classical model, can be derived from a high-order machine model. While the classical model is obtained by identifying small and large parameters in the high-order model, and setting them to zero and infinity, respectively, the library of second-order models are obtained by identifying fast and slow states in the high-order model, and replacing differential equations for the fast states with algebraic counterparts, referred to as approximate manifolds (zero-order or first-order). The library of second-order models were validated by comparing their responses to those of a high-order model, and the classical model, for given test cases.

	%%%%%%%%%%%%%%%%%%%%%%%%%%%%%%%%%%%%%%%%%%%%%%%%%%%%%%%%%%%%%%%%%%%%%%%%%%%%%%%%%%%%%%%%%%%%%%%%%%%%%%%%%%%%%%%%%%%%%%%%%%%%%%%%%%%%%%%%%%%%%%%%%%%%%%%%%%%%%%%%%%%%%%%%%%%%%%%%%%%%%%%%%%%%%%%%%%%%%%%%%%%%%%%%%%%%%%%%%%%%%%%%%%%%%%%%
			
	\begin{appendix}
		\section{Zero-Order Approximate Manifolds}
		\label{appendix}
		In this section, we present the zero-order approximate manifolds that we formulated for the fast states identified in Sec. \ref{sec:observation}. These manifolds are used in our formulation of the elemental model, the damped model, and the semi-damped model. The following zero-order approximate manifolds are common to the elemental model, the damped model, and the semi-damped model.
		\begin{equation}
		\begin{split}
		E_{f,0}(t) =&\ \frac{K_u\left(V_r^{(s)}-V^{(s)}\right)}{K_f},\quad T_{m,0}(t) = P_{u,0}(t),\\
		P_{u,0}(t) =&\ P_{c} - \bar{D}_{0}\left({{\omega}}^{(s)}(t)-\omega_0\right),\quad P_{b_1,0}(t) = 0,\\
		P_{b_2,0}(t) =&\ 0,\quad P_{a_1,0}(t) = 0,\quad P_{a_2,0}(t) = 0,\\
		U_{f,0}(t) =&\ K_fE_{f,0}(t),\quad 	\bar{U}_{f,0}(t) =\ \frac{\bar{K}_u}{\bar{\tau}_u}E_{f,0}(t),
		\end{split}\label{eqn:zeromanifold}
		\end{equation}
		where `$0$' subscripts are used to denote a zero-order approximations.

		\subsection{The Elemental Model}
		\label{appendix_elem}
		In other to formulate the elemental model, zero-order manifolds were developed by setting $\tau_{d''}$, $\tau_{d'}$ and all $\bm{\mathcal{O}}\left(\epsilon\right)$ parameters in \eqref{eqn:compact15} to zero to give \eqref{eqn:zeromanifold} and:
		\begin{equation}
		\begin{split}
		\Phi_{q,0}^{(e)}(t) =& -X^{(e)}I_{q},\\
		\Phi_{d,0}^{(e)}(t) =& -X^{(e)}I_{d},\\
		E_{q',0}(t) =& -\left(X_d-X_{d'}\right)I_{d} + E_{f,0}(t),\\
		E_{d',0}(t) =& \left(X_q-X_{q'}\right)I_{q},\\
		\Phi_{q_2,0}(t) =&-\left(X_q-X_{k}\right)I_{q},\\
		\Phi_{d_1,0}(t) =& -\left(X_d-X_{k}\right)I_{d} + E_{f,0}(t),\\
		\Phi_{q,0}(t) =& -R_{s}^{(e)}I_{d} - V^{(l)}\sin\left({\delta}^{(s)}(t)-\delta^{(l)}\right),\\
		\Phi_{d,0}(t) =&\ R_{s}^{(e)}I_{q} + V^{(l)}\cos\left({\delta}^{(s)}(t)-\delta^{(l)}\right).
		\end{split}\label{eqn:zeromanifold_1}
		\end{equation}
		The output voltage is described by: \begin{math}
		V_{q}^{(s)}=\ R^{(e)}I_{q} + X^{(e)}I_{d} + V^{(l)}\cos\left({\delta}^{(s)}(t)-\delta^{(l)}\right),\;
		V_{d}^{(s)}=\ R^{(e)}I_{d} - X^{(e)}I_{q} + V^{(l)}\sin\left({\delta}^{(s)}(t)-\delta^{(l)}\right), \end{math} and the output current is described by \begin{math}I_{q} = \frac{R_{s}^{(e)}\left(\frac{K_u\left(V_r^{(s)}-V^{(s)}\right)}{K_f}\right)}{\left(R_{s}^{(e)}\right)^2+X_q^{(e)}X_d^{(e)}}-\frac{R_{s}^{(e)}\left(V^{(l)}\cos\left({\delta}^{(s)}(t)-\delta^{(l)}\right)\right)}{\left(R_{s}^{(e)}\right)^2+X_q^{(e)}X_d^{(e)}}+\frac{X_d^{(e)} \left(V^{(l)}\sin\left({\delta}^{(s)}(t)-\delta^{(l)}\right)\right)}{\left(R_{s}^{(e)}\right)^2+X_q^{(e)}X_d^{(e)}},\;
		I_{d} = \frac{X_q^{(e)}\left(\frac{K_u\left(V_r^{(s)}-V^{(s)}\right)}{K_f}\right)}{\left(R_{s}^{(e)}\right)^2+X_q^{(e)}X_d^{(e)}}-\frac{X_q^{(e)}\left(V^{(l)}\cos\left({\delta}^{(s)}(t)-\delta^{(l)}\right)\right)}{\left(R_{s}^{(e)}\right)^2+X_q^{(e)}X_d^{(e)}}-\frac{R_{s}^{(e)}\left(V^{(l)}\sin\left({\delta}^{(s)}(t)-\delta^{(l)}\right)\right)}{\left(R_{s}^{(e)}\right)^2+X_q^{(e)}X_d^{(e)}}.
		\end{math} 
		
		\subsection{The Damped Model}
		\label{appendix_damp}
		In other to formulate the damped model, the following zero-order manifolds were developed by setting $R_s=0$, $R^{(e)}=0$, $\frac{1}{\omega_0}=0$ and $\frac{\omega^{(s)}(t)}{\omega_0}=1$:
		\begin{equation}
		\begin{split}
		\Phi_{q,0}(t) =& - V^{(l)}\sin\left({\delta}^{(s)}(t)-\delta^{(l)}\right),\\
		\Phi_{d,0}(t) =&\ V^{(l)}\cos\left({\delta}^{(s)}(t)-\delta^{(l)}\right),\\
		\Phi_{q,0}^{(e)}(t) =&\ V_{d}^{(s)} - V^{(l)}\sin\left({\delta}^{(s)}(t)-\delta^{(l)}\right),\\
		\Phi_{d,0}^{(e)}(t) =&- V_{q}^{(s)} + V^{(l)}\cos\left({\delta}^{(s)}(t)-\delta^{(l)}\right),
		\end{split}\label{eqn:PhiZeroMan}
		\end{equation} from where it follows that: \begin{math}
		I_{q} = \frac{\left(X_{q'}-X_{q''}\right)}{\left(X_{q'}-X_{k}\right)\left(X_{q''}^{(e)}\right)}\Phi_{q_2}(t)-\frac{\left(X_{q''}-X_{k}\right)}{\left(X_{q'}-X_{k}\right)\left(X_{q''}^{(e)}\right)}E_{d'}(t)+\frac{V^{(l)}\sin\left({\delta}^{(s)}(t)-\delta^{(l)}\right)}{X_{q''}^{(e)}},\end{math} and \begin{math}
		I_{d} =\ \frac{\left(X_{d'}-X_{d''}\right)}{\left(X_{d'}-X_{k}\right)\left(X_{d''}^{(e)}\right)}\Phi_{d_1}(t)+\frac{\left(X_{d''}-X_{k}\right)}{\left(X_{d'}-X_{k}\right)\left(X_{d''}^{(e)}\right)}E_{q'}(t)-\frac{V^{(l)}\cos\left({\delta}^{(s)}(t)-\delta^{(l)}\right)}{X_{d''}^{(e)}}.
		\end{math}
		
		A second set of zero-order manifolds was developed by setting $\tau_{d'}$, and all $\bm{\mathcal{O}}\left(\epsilon\right)$ parameters except $\tau_{q''}$ and $\tau_{q'}$, to zero to give \eqref{eqn:zeromanifold} and:
		\begin{equation}
		\begin{split}
		E_{q',0}(t) =&\ \frac{X_{d'}^{(e)}}{X_{d}^{(e)}}E_{f,0}(t)-\frac{N_{d}}{D_{d}}\dot{V}_q^{(l)}\\&+\frac{X_d-X_{d'}}{X_d^{(e)}}V^{(l)}\cos\left({\delta}^{(s)}(t)-\delta^{(l)}\right),\\
		\end{split}\label{eqn:zeromanifold2}
		\end{equation}where \begin{math}
		N_{d} = \tau_{d'}\tau_{d''}X_{d'}^{(e)}X_{k}^{(e)}(X_{d}-X_{d'})(X_{d'}-X_{d''})(X_{d'}-X_{k}), \text{ and } D_{d} = \tau_{d'}X_{d}^{(e)}(X_{d'}^{(e)})^2(X_{d'}-X_{k})^2-\tau_{d''}X_{d}^{(e)}(X_{k}^{(e)})^2(X_{d}-X_{d'})(X_{d'}-X_{d''}).
		\end{math}
		
		\subsection{The Semi-Damped Model}
		\label{appendix_semidamp}
		In other to formulate the semi-damped model, the following zero-order manifolds were developed by setting $R_s=0$, $R^{(e)}=0$, $\frac{1}{\omega_0}=0$, $\frac{\omega^{(s)}(t)}{\omega_0}=1$, $\tau_{q''}=0$ and $\tau_{d''}=0$:
		\begin{equation}
		\begin{split}
		\Phi_{q,0}(t) =& - V^{(l)}\sin\left({\delta}^{(s)}(t)-\delta^{(l)}\right),\\
		\Phi_{d,0}(t) =&\ V^{(l)}\cos\left({\delta}^{(s)}(t)-\delta^{(l)}\right),\\
		\Phi_{q,0}^{(e)}(t) =&\ V_{d}^{(s)} - V^{(l)}\sin\left({\delta}^{(s)}(t)-\delta^{(l)}\right),\\
		\Phi_{d,0}^{(e)}(t) =&- V_{q}^{(s)} + V^{(l)}\cos\left({\delta}^{(s)}(t)-\delta^{(l)}\right),\\
		\Phi_{q_2,0}(t) =&-\left(X_{q'}-X_{k}\right)I_{q}-E_{d'}(t),\\
		\Phi_{d_1,0}(t) =& -\left(X_{d'}-X_{k}\right)I_{d} + E_{q'}(t),
		\end{split}\label{eqn:PhiZeroMan1}
		\end{equation}
		from where it follows that: \begin{math}
		I_{q} =-\frac{1}{X_{q'}^{(e)}}E_{d'}(t)+\frac{V^{(l)}\sin\left({\delta}^{(s)}(t)-\delta^{(l)}\right)}{X_{q'}^{(e)}},\;
		I_{d} = \frac{1}{X_{d'}^{(e)}}E_{q'}(t)-\frac{V^{(l)}\cos\left({\delta}^{(s)}(t)-\delta^{(l)}\right)}{X_{d'}^{(e)}}.
		\end{math}
		
		A second set of zero-order manifolds was developed by setting $\tau_{d''}$, $\tau_{d'}$ and all $\bm{\mathcal{O}}\left(\epsilon\right)$ parameters except $\tau_{q'}$ to zero to give \eqref{eqn:zeromanifold} and:
		\begin{equation}
		\begin{split}
		E_{q',0}(t) =&\ \frac{X_{d'}^{(e)}}{X_{d}^{(e)}}E_{f,0}(t)+\frac{X_d-X_{d'}}{X_d^{(e)}}V^{(l)}\cos\left({\delta}^{(s)}(t)-\delta^{(l)}\right),
		\end{split}\label{eqn:zeromanifold3}
		\end{equation}where \begin{math}
		N_{d} = \tau_{d'}\tau_{d''}X_{d'}^{(e)}X_{k}^{(e)}(X_{d}-X_{d'})(X_{d'}-X_{d''})(X_{d'}-X_{k}),\text{ and }D_{d} = \tau_{d'}X_{d}^{(e)}(X_{d'}^{(e)})^2(X_{d'}-X_{k})^2-\tau_{d''}X_{d}^{(e)}(X_{k}^{(e)})^2(X_{d}-X_{d'})(X_{d'}-X_{d''}).
		\end{math}
		
	\end{appendix}

	\bibliographystyle{IEEEtran}
	% argument is your BibTeX string definitions and bibliography database(s)
	\bibliography{TPSrefs}

% Generated by IEEEtran.bst, version: 1.14 (2015/08/26)
\begin{thebibliography}{10}
\providecommand{\url}[1]{#1}
\csname url@samestyle\endcsname
\providecommand{\newblock}{\relax}
\providecommand{\bibinfo}[2]{#2}
\providecommand{\BIBentrySTDinterwordspacing}{\spaceskip=0pt\relax}
\providecommand{\BIBentryALTinterwordstretchfactor}{4}
\providecommand{\BIBentryALTinterwordspacing}{\spaceskip=\fontdimen2\font plus
\BIBentryALTinterwordstretchfactor\fontdimen3\font minus
  \fontdimen4\font\relax}
\providecommand{\BIBforeignlanguage}[2]{{%
\expandafter\ifx\csname l@#1\endcsname\relax
\typeout{** WARNING: IEEEtran.bst: No hyphenation pattern has been}%
\typeout{** loaded for the language `#1'. Using the pattern for}%
\typeout{** the default language instead.}%
\else
\language=\csname l@#1\endcsname
\fi
#2}}
\providecommand{\BIBdecl}{\relax}
\BIBdecl

\bibitem{Kundur1994}
P.~Kundur, N.~J. Balu, and M.~G. Lauby, \emph{Power system stability and
  control}.\hskip 1em plus 0.5em minus 0.4em\relax McGraw-Hill, 1994.

\bibitem{sauer2006power}
P.~Sauer and A.~Pai, \emph{Power System Dynamics and Stability}.\hskip 1em plus
  0.5em minus 0.4em\relax Stipes Publishing L.L.C., 2006.

\bibitem{KrauseWasynczuk2013}
P.~Krause, O.~Wasynczuk, S.~Sudhoff, and S.~Pekarek, \emph{Analysis of Electric
  Machinery and Drive Systems}, ser. IEEE Press Series on Power
  Engineering.\hskip 1em plus 0.5em minus 0.4em\relax Wiley, 2013.

\bibitem{Wang2007}
L.~Wang, J.~Jatskevich, and H.~W. Dommel, ``Re-examination of synchronous
  machine modeling techniques for electromagnetic transient simulations,''
  \emph{IEEE Transactions on Power Systems}, vol.~22, no.~3, pp. 1221--1230,
  Aug. 2007.

\bibitem{Crary1947}
S.~Crary, \emph{Power System Stability: Transient stability}, ser. General
  Electric series.\hskip 1em plus 0.5em minus 0.4em\relax John Wiley, 1947.

\bibitem{Kimbark1956}
E.~Kimbark, \emph{Power Systems Stability. Vol. 3. Synchronous Machines}.\hskip
  1em plus 0.5em minus 0.4em\relax Wiley, 1956.

\bibitem{pai1989energy}
A.~Pai, \emph{Energy Function Analysis for Power System Stability}, ser. Power
  Electronics and Power Systems.\hskip 1em plus 0.5em minus 0.4em\relax
  Springer, Boston, MA, 1989.

\bibitem{anderson2003power}
P.~M. Anderson and A.~A. Fouad, \emph{Power system control and stability}, ser.
  IEEE Press power engineering series.\hskip 1em plus 0.5em minus 0.4em\relax
  IEEE Press, 2003.

\bibitem{classicalmodel2015}
S.~Y. Caliskan and P.~Tabuada, ``Uses and abuses of the swing equation model,''
  in \emph{Proc. of IEEE Conference on Decision and Control (CDC)}, Dec 2015,
  pp. 6662--6667.

\bibitem{Weckesser2013}
T.~Weckesser, H.~Jóhannsson, and J.~Østergaard, ``Impact of model detail of
  synchronous machines on real-time transient stability assessment,'' in
  \emph{Proc. of the IREP Symposium Bulk Power System Dynamics and Control - IX
  Optimization, Security and Control of the Emerging Power Grid}, Aug 2013, pp.
  1--9.

\bibitem{kokotovicSingular}
P.~Kokotovi{\'c}, H.~K. Khalil, and J.~O'Reilly, \emph{Singular Perturbation
  Methods in Control: Analysis and Design}, ser. Classics in Applied
  Mathematics.\hskip 1em plus 0.5em minus 0.4em\relax Society for Industrial
  and Applied Mathematics, 1986.

\bibitem{Khalil2013}
H.~K. Khalil, \emph{Nonlinear Systems}.\hskip 1em plus 0.5em minus 0.4em\relax
  Pearson Education, Limited, 2013.

\bibitem{joechow1982}
J.~H. Chow, \emph{Time-Scale Modeling of Dynamic Networks with Applications to
  Power Systems}, B.~A.V. and T.~M., Eds.\hskip 1em plus 0.5em minus
  0.4em\relax Springer, 1982.

\bibitem{SauerAhmedKokotovic1988}
P.~W. Sauer, S.~Ahmed-Zaid, and P.~V. Kokotovic, ``An integral manifold
  approach to reduced order dynamic modeling of synchronous machines,''
  \emph{IEEE Transactions on Power Systems}, vol.~3, no.~1, pp. 17--23, Feb.
  1988.

\bibitem{SauerKokotovic1989}
P.~V. Kokotovic and P.~W. Sauer, ``Integral manifold as a tool for
  reduced-order modeling of nonlinear systems: A synchronous machine case
  study,'' \emph{IEEE Transactions on Circuits and Systems}, vol.~36, no.~3,
  pp. 403--410, Mar. 1989.

\bibitem{IEEE_ExcMod}
``{IEEE Recommended Practice for Excitation System Models for Power System
  Stability Studies},'' \emph{{IEEE Std} 421.5-2016 (Revision of {IEEE Std}
  421.5-2005)}, pp. 1--207, Aug. 2016.

\bibitem{Degov1MOD}
\BIBentryALTinterwordspacing
{PowerWorld corporation}. (2017) Woodward diesel governor model. [Online].
  Available: \url{{https://www.powerworld.com}}
\BIBentrySTDinterwordspacing

\bibitem{fouad1992power}
A.~Fouad and V.~Vittal, \emph{Power System Transient Stability Analysis Using
  the Transient Energy Function Method}.\hskip 1em plus 0.5em minus 0.4em\relax
  Prentice Hall, 1992.

\end{thebibliography}
\end{document}